\documentclass[twocolumn,superscriptaddress,floatfix,prb]{revtex4}
\usepackage{graphicx,amsfonts,amssymb,amsmath,hyperref,hypcap,enumerate}
\usepackage[mathscr]{euscript}
\usepackage{color}

\newcommand{\hatk}{\hat{k}}
\newcommand{\qq}{\boldsymbol{q}}
\newcommand{\pp}{\boldsymbol{p}}
\newcommand{\kk}{\boldsymbol{k}}
\newcommand{\DD}{\boldsymbol{D}}
\newcommand{\dd}{\boldsymbol{d}}
\newcommand{\bs}{\boldsymbol{s}}

\begin{document}
\title{Nematic and chiral superconductivity induced by odd-parity fluctuations}

\author{Fengcheng Wu}
\email{fengcheng.wu@anl.gov}
\affiliation{Materials Science Division, Argonne National Laboratory, Argonne, Illinois 60439, USA}

\author{Ivar Martin}
\affiliation{Materials Science Division, Argonne National Laboratory, Argonne, Illinois 60439, USA}

\date{\today}

\begin{abstract}
Recent experiments indicate that superconductivity in Bi$_2$Se$_3$ intercalated with Cu, Nb or Sr is nematic with rotational symmetry breaking. Motivated by this observation, we present a model study of nematic and chiral superconductivity induced by odd-parity fluctuations. We show that odd-parity fluctuations in the two-component $E_u$ representation of $D_{3d}$ crystal point group can generate attractive interaction in both the even-parity $s$-wave and odd-parity $E_u$ pairing channels, but repulsive interaction in other odd-parity pairing channels. Coulomb repulsion can suppress $s$-wave pairing relative to $E_u$ pairing, and thus the latter can have a higher critical temperature. $E_u$ pairing has two distinct phases: a nematic phase and a chiral phase, both of which can be realized in our model. When $s$-wave and $E_u$ pairings have similar instability temperature, we find an intermediate phase in which both types of pairing coexist.
\end{abstract}

\maketitle

\section{Introduction}
The theoretical identification of time-reversal invariant topological insulators\cite{KaneMele,FuKaneMele} has sparked a great discovery of topological states in various forms of matter, including insulators\cite{HasanKane, QiZhang}, superconductors\cite{QiZhang} and semimetals\cite{Liu2014, Xu2015}. A topological superconductor is enriched by its intrinsic particle-hole symmetry, which protects zero-energy Majorana modes on boundaries and in vortices\cite{QiZhang}. Topological superconductivity is being actively studied in both theory\cite{Fu_Kane, Qi2009, Sau2010, Lutchyn2010} and experiment\cite{nadj2014,albrecht2016}.

Recent experiments have identified Bi$_2$Se$_3$ intercalated with Cu, Nb or Sr as a candidate system for topological superconductor. 
Many bulk properties in the superconducting state of doped Bi$_2$Se$_3$ display a uniaxial anisotropy in response to an in-plane magnetic field, which include Knight shift\cite{matano2016spin}, upper critical field\cite{yonezawa2017,pan2016rotational}, magnetic torque\cite{Asaba2017} and specific heat\cite{yonezawa2017}. Therefore, the superconducting state breaks the lattice discrete rotational symmetry, and can be termed as {\em nematic}. Specific heat\cite{Ando2011} and penetration depth measurement\cite{Smylie2016} have shown the absence of line nodes in the superconducting state. Given these experimental observations, the nematic state is most consistent with an $E_u$ pairing channel that has two components and odd-parity symmetry\cite{Fu_Nematic}. Here $E_u$ is one of the symmetry representations allowed by the $D_{3d}$ point group of Bi$_2$Se$_3$. The odd-parity nematic state can be a fully-gapped time-reversal-invariant topological superconductor\cite{Fu_Nematic}. So far, experimental evidence of surface Majorana states associated with topological superconductivity has been not conclusive\cite{Sasaki2011, Levy2013}. On the theoretical side, different aspects of the nematic states have been explored, including bulk properties\cite{hashimoto2013bulk,Nagai2016,Venderbos2016}, surface states\cite{Zhang_2014}, vortex states\cite{Wu2017, zyuzin2017}, and the interplay between $E_u$ superconductivity and magnetism\cite{Law2017,Chirolli2017}.

The basic question, which remains largely open\cite{Fu2016, Behnia2017}, is the underlying microscopic mechanism for the odd-parity nematic superconductivity in doped Bi$_2$Se$_3$. In the pioneering work of Fu and Berg\citep{Fu_Berg}, they demonstrated that pairing instability in the odd-parity channels can be generated by a simple type of attractive interaction in doped Bi$_2$Se$_3$. However, the odd-parity $A_{1u}$ pairing channel has a higher critical temperature than the $E_u$ pairing channel in their model. 

Odd-parity pairing can be induced by magnetic fluctuations, as in the case of superfluid Helium-3\cite{Legget1975} and in strongly correlated materials like Sr$_2$RuO$_4$\cite{RiceSigrist} and UPt$_3$\cite{Nomoto2016}.  It has recently been proposed that odd-parity pairing can also be induced by odd-parity fluctuations in a system with strong spin-orbit coupling, time reversal and inversion symmetries\cite{Kozii2015, Wang2016, Ruhman2017}. As doped Bi$_2$Se$_3$ is likely a weakly correlated material, we study superconductivity induced by odd-parity fluctuations in this paper.

In Ref.~\onlinecite{Kozii2015}, Kozii and Fu have studied the most symmetric group $O(3)$ in three dimensions, and found that odd-parity fluctuation in pseudoscalar and vector representations generate attractive interaction in both conventional even-parity $s$-wave pairing channel and odd-parity pairing channels, while fluctuation in the multipolar channel only generates attractive interaction in the $s$-wave channel. 
Our work builds upon Ref.~\onlinecite{Kozii2015}. We apply a similar approach to doped Bi$_2$Se$_3$ which has a $D_{3d}$ point group symmetry.
Symmetry classifications of odd-parity fluctuations for $O(3)$ and $D_{3d}$ groups are different.
Our main results can be summarized as follows. Odd-parity fluctuations in the $E_u$ representation of the $D_{3d}$ point group can induce {\em attractive} interaction in both the $s$-wave and odd-parity $E_u$ pairing channels, but {\em repulsive} interaction in the other two odd-parity $A_{1u}$ and $A_{2u}$ pairing channels.  
The competition between $s$-wave and $E_u$ pairings can be further tuned by Coulomb repulsion, which has the strongest pair-breaking effect in the $s$-wave channel. 

The organization of this paper is the following. In Sec.~\ref{FlucSuc}, we study odd-parity fluctuations and superconductivity. The fluctuations are possibly induced by electron-phonon interaction. We use an approach that closely follow that in Ref.~\onlinecite{Kozii2015}. Essential details of the approach will be presented to make the discussion self-contained. We obtain a phase diagram (Fig.~\ref{Fig:U_phase_diagram}) as a function of phenomenological parameters $\gamma_i$ (i=1,2,3)  and $U$. $\gamma_i$, introduced in Eq.~(\ref{gammaA}), parametrize odd-parity particle-hole fluctuations in $E_u$ representation. $U$ is the repulsive interaction in the $s$-wave pairing channel, which can arise from Coulomb repulsion. There is a critical $U_c$, above which $E_u$ pairing has a higher critical temperature compared to the $s$-wave pairing.  $E_u$ superconductivity supports two distinct phases\cite{Venderbos2016_pairing}: nematic and chiral, both of which can be realized in the parameter space of $\gamma_i$.
In Sec.~\ref{coexistence}, we study a phase in the vicinity of $U_c$, where even-parity $s$-wave and odd-parity $E_u$ pairing can coexist. The coexistence phase spontaneously breaks both time-reversal and lattice discrete rotational symmetries. The gap structures in different superconductivity phases are reviewed.
In Sec.~\ref{discussion}, we discuss our work in the context of previous studies.
We present some related materials in the Appendixes. Appendix~\ref{appA} shows that an on-site repulsion in Bi$_2$Se$_3$ generates repulsive interaction in both $s$-wave and $A_{2u}$ pairing channels, but not in the $E_u$ channel. In Appendixes~\ref{appB} and \ref{appC}, we show that odd-parity fluctuations in $A_{1u}$ ($A_{2u}$) representation can generate $A_{1u}$ ($A_{2u}$) Cooper pairing besides the usual $s$-wave pairing.

Before ending the Introduction section, we mention that odd-parity particle-hole fluctuations can become unstable and lead to spontaneous parity-breaking phases\cite{Fu2015}, which have been recently observed in Cd$_2$Re$_2$O$_7$\cite{Harter2017}.

\section{Two-component Odd-parity fluctuation and Superconductivity}
\label{FlucSuc}
Electronic bands in Bi$_2$Se$_3$ are doubly degenerate at every $\kk$ point due to the presence of both time-reversal and inversion symmetries. When Bi$_2$Se$_3$ is intercalated with Cu, Nb or Sr, the chemical potential lies in the conduction bands. As attractive interaction induced by fluctuations typically occurs in a small energy window around chemical potential, we will only retain the lowest conduction bands in our theory. The Fermi surface of Bi$_2$Se$_3$ at low electron doping level is approximately spherical\cite{Hasan2010,Lawson2012}. Therefore, we approximate the conduction band by a parabolic dispersion:
\begin{equation}
H_0=\sum_{\kk}(\frac{\hbar \kk^2}{2m}-\mu) c_{\kk}^{\dagger} c_{\kk},
\label{HH0}
\end{equation}
which is intended to describe physics near the chemical potential $\mu$. 
$c_{\kk}^\dagger$ represents a two-component spinor $(c_{\kk \uparrow}^\dagger,c_{\kk \downarrow}^\dagger)$, which is understood to be in the ``manifestly covariant
Bloch basis"(MCBB)\cite{Fu2015}. Here $\uparrow$ and $\downarrow$ represent pseudospin instead of real spin because of strong spin-orbit coupling. Nevertheless, the pseudospin in the MCBB transforms in the same way as the real spin of a free electron under symmetry operations. In particular, the transformations under time reversal ($\hat{\mathcal{T}}$) and inversion ($\hat{\mathcal{P}}$) operations are:
\begin{equation}
\hat{\mathcal{T}}c_{\kk \alpha}^\dagger\hat{\mathcal{T}}^{-1} = \epsilon_{\alpha \beta} c_{-\kk \beta}^\dagger,\,\,\,\,
\hat{\mathcal{P}}c_{\kk \alpha}^\dagger\hat{\mathcal{P}}^{-1} = c_{-\kk \alpha}^\dagger,
\end{equation}
where $\epsilon_{\alpha \beta}$ is the fully antisymmetric tensor with $\epsilon_{\uparrow \downarrow}=1$.

To study electron-phonon interaction, we focus on phonons at the Brillouin zone center, which can be classified by the $D_{3d}$ point group of Bi$_2$Se$_3$. To be specific, we consider $E_u$ phonons that are {\em odd} under inversion and have two degenerate modes. The coupling between electrons and $E_u$ phonons can be expressed as:
\begin{equation}
\begin{aligned}
H_{el-ph,0}&=\phi_x \hat{Q}_x+ \phi_y \hat{Q}_y, \\
\hat{Q}_{a}&=\frac{1}{2}\sum_{\kk} c^\dagger_{\kk} \Gamma_{a}(\kk) c_{\kk},
\end{aligned}
\label{Helph}
\end{equation}
where the Hermitian operators $(\phi_x, \phi_y)$ represent the $E_u$ phonons, and also take into account all coupling constants. $\Gamma_{x, y}(\kk)$ are $2\times 2$ matrices in the pseudospin space.
As $H_{el-ph,0}$ should be invariant under all symmetries that the system has, the operators $\hat{Q}_{x,y}$ are Hermitian, time reversal symmetric and form a two-component $E_u$ representation.  By Hermiticity, we can express $\Gamma_{x, y}(\kk)$ using identity matrix  $s_0$ and Pauli matrices $\bs$:
\begin{equation}
\Gamma_{a}(\kk)=\tilde{D}_a(\kk)s_0+\DD_a(\kk) \cdot \bs,
\label{GammaDD}
\end{equation}
where both the scalar $\tilde{D}_a$ and the vector $\DD_a$ are real.
By time reversal symmetry, we require $\tilde{D}_a(\kk)=\tilde{D}_a(-\kk)$ and $\DD_a(\kk)=-\DD_a(-\kk)$.
On the other hand, $\hat{Q}_{a}$ is odd under inversion, which leads to $\Gamma_{a}(\kk)=-\Gamma_{a}(-\kk)$. Therefore, $\tilde{D}_a(\kk)$ must {\em vanish}. 
In our low-energy theory, odd-parity phonons couple to electron's spin, which is possible due to the presence of strong spin-orbit coupling.

The form factors $\Gamma_{x,y}(\kk) $ are further restricted by other point group symmetries. There are three basis functions separately for $\Gamma_{x}$ and $\Gamma_{y}$ to first order of $\kk$ in the $E_u$ representation, as listed in Table \ref{tab:form}. In general, $\Gamma_{x,y}(\kk) $ is a linear combination of these three basis functions:
\begin{equation}
\Gamma_{a}(\kk)=\gamma_1 \Gamma_{a}^{(1)}(\kk)+\gamma_2 \Gamma_{a}^{(2}(\kk) +\gamma_3 \Gamma_{a}^{(3)}(\kk)
\label{gammaA}
\end{equation}
where $\gamma_i$ are real parameters that are not fixed by symmetries.  We will take $\gamma_i$ as free parameters and study phase diagrams in this parameter space.

\begin{table}[t]
\caption{Linear order expansion of odd-parity form factors in different symmetry representations of $D_{3d}$ point group\cite{Venderbos2016_pairing}.  $A_{1u}$ and $E_u$ representations have multiple basis functions in lowest order expansion. $\hat{k}_i$ denotes $k_i/|\kk|$.}
\begin{tabular}{c | c}
\hline
\hline
Symmetry  & Form factors \\
\hline  
$A_{1u}$ & $ \Gamma_1^{(1)}= \frac{1}{\sqrt{2}}(\hatk_x s_x+\hat{k}_y s_y)$, $\Gamma_1^{(2)}=\hatk_z s_z$   \\
\hline
$A_{2u}$ & $\Gamma_2^{(1)} = \frac{1}{\sqrt{2}}(\hatk_x s_y - \hatk_y s_x) $   \\
\hline
$E_u$ & $\begin{matrix} \Gamma_x^{(1)} = \hatk_x s_z , \Gamma_x^{(2)} = \hatk_z s_x, \Gamma_x^{(3)} =\frac{1}{\sqrt{2}}(\hatk_x s_y + \hatk_y s_x)  \\ 
                        \Gamma_y^{(1)} = \hatk_y s_z , \Gamma_y^{(2)} = \hatk_z s_y, \Gamma_y^{(3)} =\frac{1}{\sqrt{2}}(\hatk_x s_x - \hatk_y s_y)
          \end{matrix}  $\\
\hline
\hline
\end{tabular}
\label{tab:form}
\end{table}

$H_{el-ph,0}$ describes the coupling between electrons and zone-center phonon modes. We generalize the coupling to include phonon modes at  finite momentum: 
\begin{equation}
\begin{aligned}
H_{el-ph}&=\sum_{\qq}\phi_{x,\qq} \hat{Q}_x(\qq)+ \phi_{y,\qq} \hat{Q}_y(\qq), \\
\hat{Q}_{a}(\qq)&=\frac{1}{2}\sum_{\kk} c^\dagger_{\kk+\qq} [\Gamma_{a}(\kk+\qq)+\Gamma_{a}(\kk)] c_{\kk}.
\end{aligned}
\label{Helphq}
\end{equation}
In the generalization, we assume that the phonon modes vary smoothly in real space.

The electron-phonon coupling generates an effective electron-electron interaction:
\begin{equation}
H_{int}=\frac{1}{\Omega}\sum_{\qq} V_{\qq} [\hat{Q}_{x}(\qq)\hat{Q}_{x}(-\qq)+\hat{Q}_{y}(\qq)\hat{Q}_{y}(-\qq)],
\label{Hint}
\end{equation}
where $\Omega$ is the system size. 
By the definition in (\ref{Helphq}), we have $\hat{Q}_{a}(-\qq)=\hat{Q}_{a}^\dagger(\qq)$.

In $H_{int}$, we neglect the frequency dependence of  $V_{\qq}$ for simplicity. 
The point group symmetries put constraints on the momentum dependence of $V_{\qq}$:  (1) $V_{\qq}$ is an even function of $\qq$ and (2) it is invariant under a three-fold rotation of $\qq$ along $\hat{z}$ direction.

We now restrict the interaction to the Bardeen-Cooper-Schrieffer (BCS) channel:
\begin{equation}
H_{BCS}=\frac{1}{\Omega}\sum_{\kk,\kk'}V_{\alpha\beta\gamma\delta}(\kk,\kk')c^\dagger_{\kk\alpha}c^\dagger_{-\kk\beta}c_{-\kk'\gamma}c_{\kk'\delta}.
\end{equation}
The expression for the interaction vertex $V_{\alpha\beta\gamma\delta}(\kk,\kk')$ is given by:
\begin{equation}
\begin{aligned}
V_{\alpha\beta\gamma\delta}(\kk,\kk')&\\
=-\frac{1}{8}\sum_{a=x,y}\Big\{&
V_{\kk-\kk'}[\DD_a+\DD_a']\cdot \bs_{\alpha \delta}[\DD_a+\DD_a']\cdot \bs_{\beta \gamma}\\
-&V_{\kk+\kk'}[\DD_a-\DD_a']\cdot \bs_{\alpha \gamma}[\DD_a-\DD_a']\cdot \bs_{\beta \delta}
\Big\},
\end{aligned}
\end{equation}
where $\DD_a$ and $\DD_a'$ are respectively shorthand notations for $\DD_a(\kk)$ and $\DD_a(\kk')$.
Here $\DD_a(\kk)$ is the vector representation of $\Gamma_a(\kk)$, as introduced in (\ref{GammaDD}).

To minimize the number of parameters in our phenomenogical study, we further approximate $V_{\qq}$ by its value at zero momentum $V_0$.
Here $V_0<0$, representing attractive interaction induced by phonon fluctuations. Under this simplification, it is convenient to separate $V_{\alpha\beta\gamma\delta}$ to two parts: $V_{\alpha\beta\gamma\delta}=(V^e+V^o)_{\alpha\beta\gamma\delta}$. 
The expressions for $V^{e,o}$ are as follows:
\begin{equation}
\begin{aligned}
&V^e_{\alpha\beta\gamma\delta}(\kk,\kk')\\
\approx&  -\frac{V_0}{8}\sum_{a=x,y}\Big\{
(\DD_a\cdot \bs)_{\alpha \delta}(\DD_a\cdot \bs)_{\beta \gamma}-(\DD_a\cdot \bs)_{\alpha \gamma}(\DD_a\cdot \bs)_{\beta \delta}\\
+&(\DD_a'\cdot \bs)_{\alpha \delta}(\DD_a'\cdot \bs)_{\beta \gamma}-(\DD_a'\cdot \bs)_{\alpha \gamma}(\DD_a'\cdot \bs)_{\beta \delta}
\Big\}\\
=&\frac{V_0}{8}\sum_{a=x,y}(|\DD_a|^2+|\DD_a'|^2)\epsilon_{\alpha \beta} \epsilon^\dagger_{\gamma \delta},\\
&V^o_{\alpha\beta\gamma\delta}(\kk,\kk')\\
\approx&  -\frac{V_0}{8}\sum_{a=x,y}\Big\{(\DD_a\cdot \bs)_{\alpha \delta}(\DD_a'\cdot \bs)_{\beta \gamma}+(\DD_a\cdot \bs)_{\alpha \gamma}(\DD_a'\cdot \bs)_{\beta \delta}\\
&+(\DD_a'\cdot \bs)_{\alpha \delta}(\DD_a\cdot \bs)_{\beta \gamma}+(\DD_a'\cdot \bs)_{\alpha \gamma}(\DD_a\cdot \bs)_{\beta \delta}]
\Big\}\\
=&\frac{V_0}{4}\sum_{a=x,y}\Big\{[(\DD_a\cdot \bs)\epsilon]_{\alpha \beta} [(\DD_a'\cdot \bs)\epsilon]^\dagger_{\gamma \delta}\\
&\quad\quad\quad\,\,-[(\DD_a\times \bs)\epsilon]_{\alpha \beta}\cdot[(\DD_a'\times \bs)\epsilon]^\dagger_{\gamma \delta}\Big\}.
\end{aligned}
\label{Veo}
\end{equation}
Here $V^{e}$ and $V^o$ are respectively even and odd functions of $\kk$ and $\kk'$, and, therefore, generate correspondingly even and odd parity pairings.
In (\ref{Veo}), the final expressions of $V^{e,o}$ are presented in a form that is suitable for BCS decomposition.
In the following subsections \ref{subA} and \ref{subB}, we study the pairing instabilities in even and odd parity channels separately and finally compare them.

\subsection{Even-parity instability}
\label{subA}
Even-parity pairing, or typically named as $s$-wave pairing, is induced by $V^e$. As we will discuss in the subsection \ref{subB}, the effective interaction $H_{int}$ (\ref{Hint}) always generates a larger instability in $s$-wave channel compared to odd-parity channels. To study competition between even and odd parity pairings, we add a repulsive interaction to $V^e$:
\begin{equation}
\begin{aligned}
H_{e}
=&\frac{1}{\Omega}\sum_{\kk,\kk'}[V^e_{\alpha\beta\gamma\delta}(\kk,\kk')+\frac{U|V_0|}{4}\epsilon_{\alpha \beta} \epsilon^\dagger_{\gamma \delta}]c^\dagger_{\kk\alpha}c^\dagger_{-\kk\beta}c_{-\kk'\gamma}c_{\kk'\delta}\\
=&\frac{V_0}{\Omega}\sum_{\kk,\kk'}[g_0(\kk)+g_0(\kk')][\frac{1}{2}\epsilon_{\alpha\beta}c^\dagger_{\kk\alpha}c^\dagger_{-\kk\beta}][\frac{1}{2}\epsilon^{\dagger}_{\gamma \delta}c_{-\kk'\gamma}   c_{\kk'\delta}],
\end{aligned}
\end{equation}
where $U>0$ characterizes the repulsive interaction and $g_0(\kk)=(|\DD_x(\kk)|^2+|\DD_y(\kk)|^2-U)/2$. 
For reasons to become clear shortly, we make the following transformation:
\begin{equation}
\begin{aligned}
g_0(\kk)+g_0(\kk')&=\frac{1}{2\kappa}[g_+(\kk)g_+(\kk')-g_-(\kk)g_-(\kk')],\\
g_{\pm}(\kk)&=g_0(\kk)\pm\kappa,
\end{aligned}
\label{fpm}
\end{equation}
where $\kappa$ is a positive parameter. We choose $\kappa$ such that:
\begin{equation}
\langle g_+(\kk) g_-(\kk) \rangle=0,
\label{fpmOrth}
\end{equation}
where $\langle ... \rangle$ denotes an average over Fermi surface, normalized so $\langle 1 \rangle=1$.
Using (\ref{fpm}), $H_e$ can be decomposed into two channels:
\begin{equation}
\begin{aligned}
H_{e}=&\frac{V_0}{2\kappa\Omega}(S_+^\dagger S_+ -S_-^\dagger S_-),\\
S_\pm^\dagger=&\frac{1}{2}\sum_{\kk\alpha\beta}g_{\pm}(\kk)\epsilon_{\alpha\beta}c^\dagger_{\kk\alpha}c^\dagger_{-\kk\beta}.
\end{aligned}
\label{He}
\end{equation}
Because $g_+(\kk)$ and $g_-(\kk)$ are orthogonal over the Fermi surface as required by (\ref{fpmOrth}), the attractive and repulsive channels respectively represented by $S_+^\dagger$ and $S_-^\dagger$ are decoupled in the linearized gap equation. Therefore, we only consider $S_+^\dagger$ in the following. The critical temperature $T_{c,s}$ for $S_+^\dagger$ channel  is determined by its linearized gap equation:
\begin{equation}
\begin{aligned}
&|V_0|\chi_s(T_{c,s})=1,\\
&\chi_s(T)=\frac{1}{2\kappa}\langle\frac{1}{2} \text{Tr} [g_+(\kk) s_0]^2\rangle \chi_0(T).
\end{aligned}
\end{equation}
Here $\chi_0$ is the standard superconductivity susceptibility: $\chi_0(T)=N(0)\int_{-\omega_D}^{\omega_D} d\varepsilon \text{tanh}[\varepsilon/(2T)]/(2\varepsilon)$, where $N(0)$ is the density of states at the Fermi energy, $\omega_D$ is the cut off energy for attractive interaction, and $T$ is the temperature. 

\subsection{Odd-parity instability}
\label{subB}
\begin{figure}[t]
	\includegraphics[width=1\columnwidth]{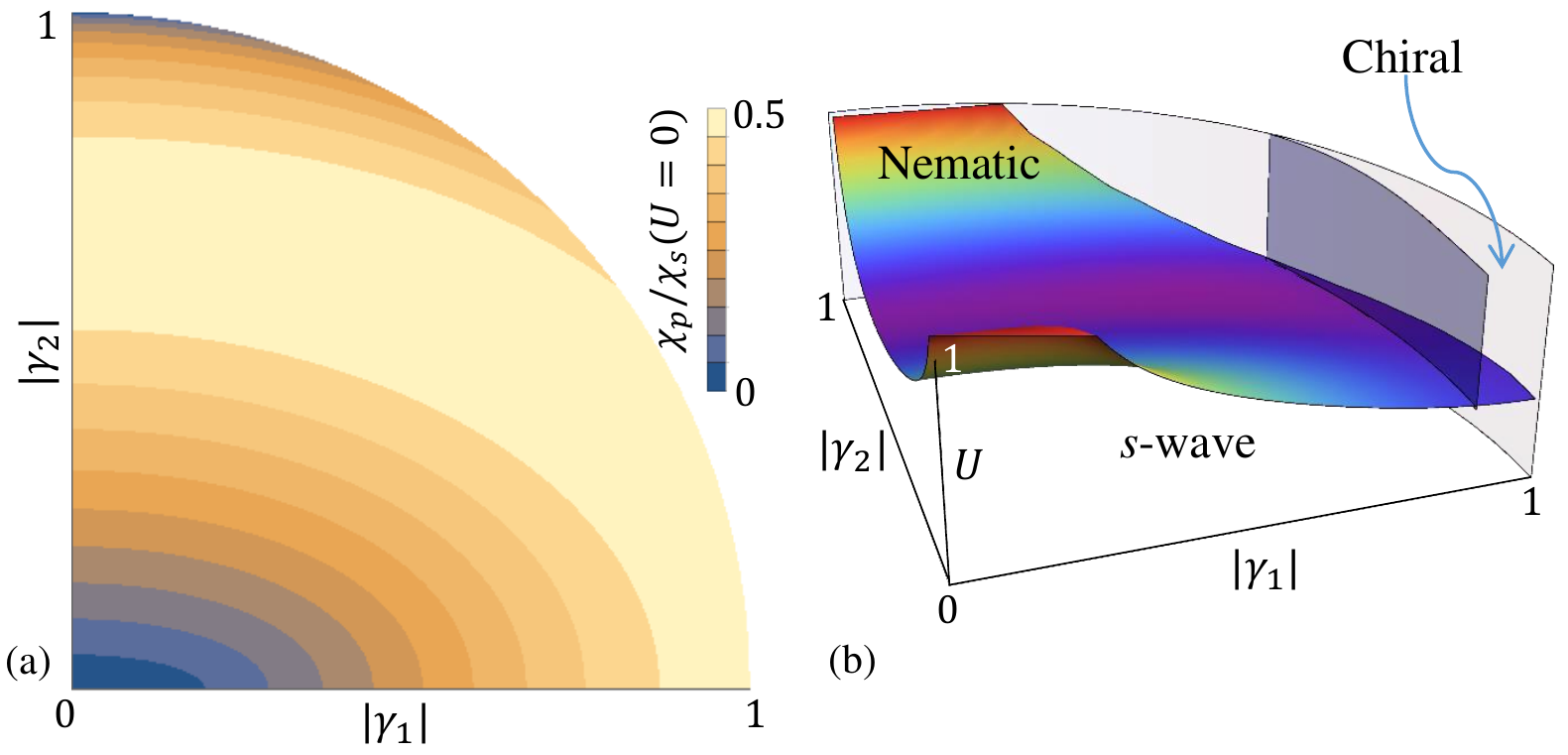}
	\caption{(a) $\chi_p(T)/\chi_s(T)$ at $U=0$ as a function of $\gamma_1$ and $\gamma_2$. (b) The surface with rainbow color represents $U_c$ at which $\chi_p(T)=\chi_s(T)$.  The odd-parity $E_u$ superconductivity supports two different phases: nematic and chiral, which are separated by the gray boundaries. In (a) and (b), we used the normalization $\gamma_1^2+\gamma_2^2+\gamma_3^2=1$ without loss of generality. Therefore, $\gamma_1^2+\gamma_2^2\leq 1$.}
	\label{Fig:U_phase_diagram}
\end{figure}

We turn to the $V^o$ interaction:
\begin{equation}
\begin{aligned}
H_{o}
=\frac{1}{\Omega}\sum_{\kk,\kk'}V^o_{\alpha\beta\gamma\delta}(\kk,\kk')c^\dagger_{\kk\alpha}c^\dagger_{-\kk\beta}c_{-\kk'\gamma}c_{\kk'\delta}.
\end{aligned}
\label{Ho}
\end{equation}
We will decompose $H_{o}$ into different odd-parity pairing channels, which are classified into different representation of the point group and generally take the form:
\begin{equation}
\hat{F}_{a}^{(i)\dagger}=\frac{1}{2}\sum_{\kk,\alpha \beta} c^\dagger_{\kk\alpha}[ \Gamma_{a}^{(i)}(\kk) \epsilon]_{\alpha \beta}c^\dagger_{-\kk\beta}.
\end{equation}
The form factor $\Gamma_a$ can be classified in the same way as those used in the particle-hole channel, which are listed in Table \ref{tab:form}.
We use subscript $a=1$ and $2$ to stand for $A_{1u}$ and $A_{2u}$ representation respectively, and $a=x$ and $y$ to denote the two components in $E_u$ representation. The superscript $i$ enumerates different basis functions within the same representation.

$H_o$ decomposed in terms of $\hat{F}_{a}^{(i)\dagger}$ has the form:
\begin{equation}
\begin{aligned}
H_{o}=&\frac{V_0}{\Omega}\Big\{-(\gamma_1 \hat{F}_{1}^{(1)}-\sqrt{2}\gamma_2 \hat{F}_{1}^{(2)} )^\dagger(\gamma_1 \hat{F}_{1}^{(1)}-\sqrt{2}\gamma_2 \hat{F}_{1}^{(2)})\\
&-\gamma_1^2 \hat{F}_{2}^{(1)\dagger}\hat{F}_{2}^{(1)}+\sum_{a=x,y}\sum_{i,j} \hat{F}_{a}^{(i)\dagger} \mathcal{W}_{ij} \hat{F}_{a}^{(j)}
\Big\},
\end{aligned}
\end{equation}
where the coefficient matrix $\mathcal{W}$ is symmetric and real:
\begin{equation}
\mathcal{W}=\begin{pmatrix}
 \gamma_1^2-\gamma_3^2 & \gamma_1\gamma_2 & 2\gamma_1\gamma_3 \\
 \gamma_1\gamma_2 & 0 & 2\gamma_2\gamma_3\\
 2\gamma_1\gamma_3 & 2\gamma_2\gamma_3  & -\gamma_1^2 
\end{pmatrix}.
\end{equation}

Because $V_0<0$, the interaction is repulsive for $A_{1u}$ and $A_{2u}$ pairing channels in $H_o$ so there is no superconductivity instability in these two channels.  

We diagonalize the matrix $\mathcal{W}$ to decompose the $E_u$ channels:
\begin{equation}
\begin{aligned}
\sum_{i,j} \hat{F}_{a}^{(i)\dagger} \mathcal{W}_{ij} \hat{F}_{a}^{(j)}
=\sum_{\nu=1}^3 w_\nu \big[\sum_i \lambda_i^{(\nu)}\hat{F}_{a}^{(i)}\big]^\dagger \big[\sum_j \lambda_j^{(\nu)}\hat{F}_{a}^{(j)}\big],
\end{aligned}
\end{equation}
where $w_\nu$ represents the $\nu$th {\em largest} eigenvalue of $\mathcal{W}$ and $(\lambda_1^{(\nu)},\lambda_2^{(\nu)},\lambda_3^{(\nu)})$ is the corresponding normalized eigenvector. We find that $w_1\geq 0$ and $w_{2, 3}\leq 0$. $w_1$ is generically positive, and it is zero only when $\gamma_{1,2}=0$ or $\gamma_{1,3}=0$. Therefore, there is generally one attractive $E_u$ pairing channel and two repulsive $E_u$ channels. Furthermore, the three $E_u$ channels are decoupled in the linearized gap equation because (1) different eigenvectors of $\mathcal{W}$ are orthogonal and (2) different form factors are orthogonal over the Fermi surface and have the same normalization for the Fermi surface average:
\begin{equation}
\langle\frac{1}{2} \text{Tr} [\Gamma_{a}^{(i)}(\kk) \Gamma_{a'}^{(i')}(\kk)]  \rangle = \frac{1}{3} \delta_{a a'}\delta_{i i'}.
\end{equation}

We focus on the attractive $E_u$ channel as summarized in the following: 
\begin{equation}
\begin{aligned}
\tilde{H}_o&=\frac{w_1 V_0}{\Omega}(\Lambda_x^\dagger \Lambda_x+\Lambda_y^\dagger \Lambda_y),\\
\Lambda_a^\dagger&=\sum_i \lambda_i^{(1)}\hat{F}_{a}^{(i)\dagger}=\frac{1}{2}\sum_{\kk,\alpha \beta} c^\dagger_{\kk\alpha}[ g_a(\kk) \epsilon]_{\alpha \beta}c^\dagger_{-\kk\beta},
\end{aligned}
\end{equation}
where we have introduced matrices $g_{x,y}$ that are defined as $g_a(\kk)=\sum_i \lambda_i^{(1)} \Gamma_a^{(i)} $. 
The corresponding linearized gap equation is:
\begin{equation}
\begin{aligned}
&|V_0|\chi_p(T_{c, p})=1,\\
&\chi_p(T)=w_1 \langle\frac{1}{2} \text{Tr} [g_x(\kk)]^2\rangle \chi_0(T)=\frac{w_1}{3} \chi_0(T),
\end{aligned}
\end{equation}
where $T_{c,p}$ is the critical temperature for the $E_u$ channel.
$\chi_p(T)$ remains the same if $g_x(\kk)$ is replaced by $g_y(\kk)$ in its expression, which is a result of the discrete lattice rotational symmetry.

As a summary, the $E_u$ phonon generates superconductivity instability in both $s$-wave channel and $E_u$ channel. We find that $\chi_p(T)$ is always weaker compared to $\chi_s(T)$ when $U=0$ (Fig.~\ref{Fig:U_phase_diagram}(a)), which means $s$-wave has higher critical temperature in this case. Nevertheless, $\chi_p(T)$ can reach about $0.5\chi_s(T)$ in a large parameter space of $\gamma_i$, indicating that the $E_u$ pairing instability can be strong. As $U$ increases, $\chi_s(T)$ decreases while $\chi_p(T)$ does not change. We can define a critical $U_c$ at which $\chi_p(T)=\chi_s(T)$.  The $s$-wave and odd-parity $E_u$ superconductivity have larger instability temperature below and above $U_c$, respectively. The phase diagram as a function of $U$ and $\gamma_i$ is summarized in Fig.~\ref{Fig:U_phase_diagram}(b).

We note that other phonon modes, which are not included in our model, generally produce attractive interaction in $s$-wave channel, but not necessarily in $E_u$ channel. Some particular phonon modes, for example $A_{2u}$ modes discussed in Appendix~\ref{appC}, can even have pair-breaking effects for $E_u$ channel. Therefore, the value of $U_c$ obtained from our model should be viewed as a lower bound of the critical repulsive interaction.

Assuming $U>U_c$, the $E_u$ superconductivity pairing is realized below $T_{c, p}$. As a two-component superconductivity, $E_u$ pairing generally has two forms: nematic and chiral. To determine which one is realized, we study the Ginzburg-Landau free energy up to fourth order in the $E_u$ pairing order parameter $(\eta_x, \eta_y)$:
\begin{equation}
\begin{aligned}
\mathcal{F}_p = &\,\,\,\,\,\,\, r_1(|\eta_x|^2+|\eta_y|^2)+b_1 (|\eta_x|^2+|\eta_y|^2)^2\\
&+b_2|\eta_x^2+\eta_y^2|^2,
\end{aligned}
\label{Fp}
\end{equation}
where the parameters $r_1$ and $b_{1,2}$ can be fully determined by the interaction $\tilde{H}_o$ under the weak-coupling analysis: 
\begin{equation}
\begin{aligned}
r_1& =\frac{1}{w_1|V_0|}(1-|V_0|\chi_p),\\
b_1& =  \langle \text{Tr} [g_x^2(\kk) g_y^2(\kk)] \rangle \beta_0,\\
b_2& =  \frac{1}{2}\langle \text{Tr} [g_x(\kk) g_y(\kk)]^2 \rangle \beta_0,
\end{aligned}
\end{equation}
where $\beta_0= 7\zeta(3)N(0)/( 16 \pi^2 T^2)$ and $\zeta(x)$ is the Riemann zeta function.
Here $b_1$ is always positive, but the sign of $b_2$ can vary as a function of $\gamma_i$. When $b_2<0$, a nematic state with real order parameter $(\eta_x, \eta_y)\propto (\cos\theta, \sin\theta)$ is favored.  Here the angle $\theta$ characterizes the nematic direction, and its value is arbitrary for the free energy $\mathcal{F}_p$ that only includes  terms  up to fourth order. For the case of $b_2>0$, a chiral state with complex order parameter $(\eta_x, \eta_y)\propto (1, \pm i)$ is favored. The nematic and chiral states respectively break the lattice rotational symmetry and time reversal symmetry. The phase boundary ($b_2=0$) between the nematic and chiral states is shown in Fig.~\ref{Fig:U_phase_diagram}(b), indicating a large parameter space in which nematic state is more favorable. It is intriguing that phononic mechanism can induce time-reversal-breaking chiral superconductivity. The competition between nematic and chiral states has been studied as a function of $\lambda_i^{(1)}$ in Ref.~\onlinecite{Venderbos2016_pairing}. Our work reveals that $\lambda_i^{(1)}$ can be derived from parameters $\gamma_i$, the latter of which could be extracted from {\em ab inito} study of electron-phonon interactions.

\section{Coexistence of even and odd parity superconductivity}
\label{coexistence}
\begin{figure}[t]
	\includegraphics[width=0.5\columnwidth]{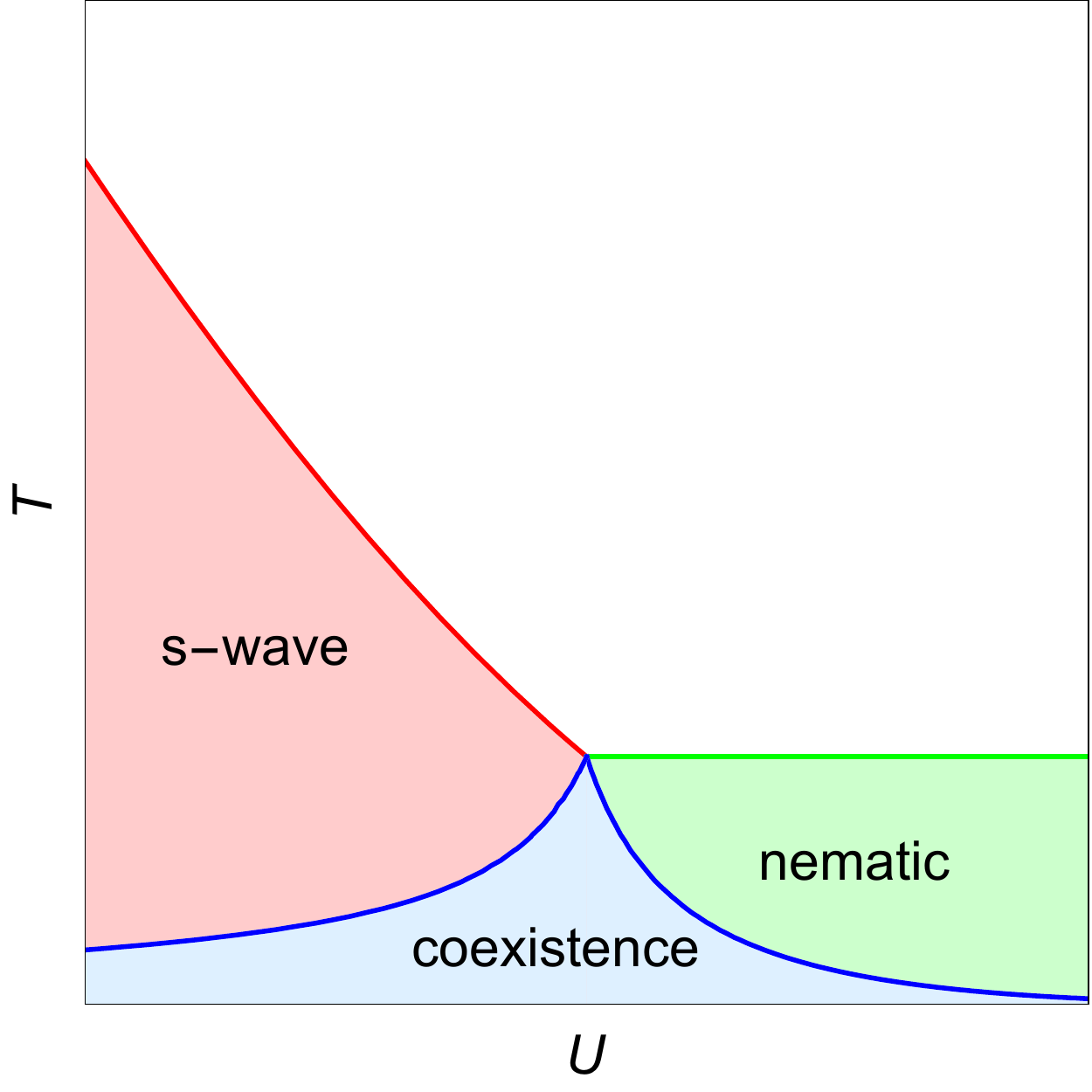}
	\caption{Schematic phase diagram as a function of repulsive interaction $U$ ($s$-wave channel) and temperature $T$. In the vicinity of $U_c$ where $s$-wave and nematic superconductivity have the same instability temperature, there is a phase where both types of superconductivity coexist with a relative phase difference $\pm \pi/2$. 
	}
	\label{Fig:U_T}
\end{figure}

At $U=U_c$, the $s$-wave and $E_u$ channel have the same critical temperature $T_{c,s}=T_{c,p}=T_c^*$.
To pin down the nature of the superconductivity below $T_c^*$, we study the Ginzburg-Landau free energy that includes both $s$-wave and $E_u$ pairing order parameters:
\begin{equation}
\begin{aligned}
\mathcal{F}= &\mathcal{F}_s+\mathcal{F}_p+\mathcal{F}_{sp}\\
\mathcal{F}_s = & r_0 |\eta_s|^2+b_0 |\eta_s|^4\\
\mathcal{F}_{sp} = & b_3\big\{ 4(|\eta_x|^2+|\eta_y|^2)|\eta_s|^2 \\& \quad+ [(\eta_x^2+\eta_y^2) \eta_s^{*2} +c.c.] \big\},
\end{aligned}
\end{equation}
where $\mathcal{F}_s$ is the free energy for $s$-wave pairing characterized by the order parameter $\eta_s$, $\mathcal{F}_p$ is give in (\ref{Fp})  and $\mathcal{F}_{sp}$ describes the coupling between $s$-wave and $E_u$ pairings. Parameters in the free energy are again obtained using weak-coupling analysis: $r_0=2\kappa(1-|V_0|\chi_s)/|V_0|$, $b_0=\frac{1}{2}\langle \text{Tr} [g_0(\kk) s_0]^4 \rangle \beta_0$
and $b_3=\frac{1}{2}\langle \text{Tr} [g_x^2(\kk) g_0^2(\kk)] \rangle \beta_0$.
Here $b_0$ and $b_3$ are always positive.

To minimize $\mathcal{F}$ below $T_c^*$ at $U=U_c$, it is most instructive to consider the case $b_2<0$.
$\mathcal{F}$ is then minimized by a state where the $s$-wave and nematic superconductivity coexist and  have a relative phase difference $\pm \pi/2$, i.e. $\eta_s=\pm i |\eta_s|$ and $(\eta_x, \eta_y)=|\eta_p|(\cos\theta, \sin\theta)$.
$|\eta_s|$ and $|\eta_p|$ are given by:
\begin{equation}
\begin{aligned}
|\eta_s|^2&=\frac{-r_0 (b_1+b_2)+r_1 b_3}{2[b_0(b_1+b_2)-b_3^2]},\\
|\eta_p|^2&=\frac{-r_1 b_0+r_0 b_3}{2[b_0(b_1+b_2)-b_3^2]}.
\end{aligned}
\label{etasetap}
\end{equation}
The coexistence of the two superconductivity order parameters requires the expressions for $|\eta_s|^2$ and $|\eta_p|^2$ in (\ref{etasetap}) to be positive, which we find to be generally satisfied in the $\gamma_i$ parameter space.

When $U$ is away from $U_c$, the coexistence state can still develop, but at a temperature lower than $T_{c,s}$ when $U<U_c$ or $T_{c,p}$ when $U>U_c$.
The schematic phase diagram as a function of $U$ and $T$ is shown in Fig.~\ref{Fig:U_T}. This coexistence phase not only breaks lattice discrete rotational symmetry because of the presence of nematic order parameter, but also breaks time reversal symmetry because of the relative phase difference $\pm\pi/2$ between the even and odd parity order parameters.

In the case of $b_2>0$, there can also be an intermediate phase between $s$-wave and chiral phases in the vicinity of $U_c$. This intermediate phase is characterized by non-zero order parameters $(\eta_s, \eta_+, \eta_-)$, where $\eta_\pm=\eta_x\pm i \eta_y$. $|\eta_+|$ and $|\eta_-|$ are generally unequal so both time reversal and discrete rotational symmetries are also broken.  

We now discuss gap structures in different phases. In the $s$-wave phase, the superconductivity gap is proportional to $g_+(\kk)$ on the Fermi surface, which is fully gapped for weak repulsion $U$. 

To study gap structure in the nematic phase, we express $g_{a}(\kk)$ for $a=x$ and $y$ in terms of a vector:
\begin{equation}
g_{a}(\kk) = \dd_{a}(\kk)\cdot \bs.
\end{equation} 
For order parameter $(\eta_x, \eta_y)$ given by $|\eta_p|(\cos\theta,\sin\theta)$, the gap is proportional to $|\dd|$ on the Fermi surface, where the vector $\dd$ is defined as $\cos\theta \dd_x +\sin\theta \dd_y $. $|\dd|$ is finite everywhere on the Fermi surface unless $\theta=n \pi/3$ (integer $n$ takes value from 0 to 5). The nematic phase realizes a fully gapped topological superconductor when $\theta \neq n \pi/3$, as it has odd parity pairing and the Fermi surface encloses only one time reversal invariant momentum\cite{Fu_Berg}. A hallmark of a time-reversal invariant topological superconductor is that it supports Majorana modes bound to surfaces and time-reversal-invariant vortex defects\cite{Qi2009, Wu2017}. When $\theta=n \pi/3$, the nematic pairing preserves one of the mirror symmetries and the gap vanishes at two opposite momenta located on the corresponding mirror-invariant plane in the Brillouin zone\cite{Fu_Nematic}. Therefore, the nematic phase with $\theta=n \pi/3$ realizes a topological Dirac superconductor with Dirac point nodes in the bulk and Majorana arcs on certain surfaces\cite{Zhang_2014}.

In the coexistence phase where $s$-wave and nematic order parameter has a phase difference $\pm \pi/2$, the Bogoliubov-de Gennes Hamiltonian is:
\begin{equation}
\mathcal{H}(\kk)=\varepsilon_0(\kk)\tau_z+|\eta_p|[\dd(\kk)\cdot\bs] \tau_x + |\eta_s| g_+(\kk) \tau_y,
\label{BdGH}
\end{equation}
which is expressed in the basis $(c_{\kk \uparrow}^\dagger,c_{\kk \downarrow}^\dagger, c_{-\kk \downarrow},-c_{-\kk \uparrow}  )$. $\varepsilon_0(\kk)=\hbar^2 \kk^2/(2m)-\mu$, and $\tau_{x, y, z}$ are Pauli matrices in the Nambu space. Here $|\eta_p|$ and $|\eta_s|$ are respectively coupled to $\tau_x$ and $\tau_y$, reflecting the $\pi/2$ phase difference. The energy spectrum of $\mathcal{H}(\kk)$ is $\pm\sqrt{\varepsilon_0(\kk)^2+|\eta_p|^2|\dd(\kk)|^2+|\eta_s|^2g_+(\kk)^2}$, which is fully gapped for any value of $\theta$. The surface Majorana zero modes presented in the nematic phase also become gapped in the coexistence phase because of broken time reversal symmetry. Such a state represents a superconducting analog of an axion insulator\cite{Qi2008}, and can have thermal Hall effect on the surface. 
Similar  phase with coexistence of  even and odd parity pairing  have been studied in Ref.~\onlinecite{Roy2014} and recently in Ref.~\onlinecite{Wang_Fu}.
A distinct feature of the coexistence phase that we obtain is that it spontaneously breaks discrete rotational symmetry besides time reversal symmetry. We also note an additional symmetry breaking in the coexistence phase. In (\ref{BdGH}), $\mathcal{H}(\kk)$ satisfies an inversion symmetry $\mathcal{H}(\kk)=\mathcal{H}(-\kk)$ when $|\eta_p|=0$, or an inversion-gauge symmetry $\tau_z\mathcal{H}(\kk)\tau_z=\mathcal{H}(-\kk)$ when $|\eta_s|=0$. In the coexistence phase, neither the inversion nor inversion-gauge symmetry remains.

The chiral phase characterized by $(\eta_x, \eta_y) \propto (1, \pm i)$ realizes a topological Weyl superconductor with bulk Weyl point nodes. The nodal structure has been extensively discussed in Refs.~\onlinecite{Venderbos2016_pairing} and \onlinecite{kozii2016}. When there is some mixing between $s$-wave and chiral superconductivity near $U_c$, the Weyl points remain robust unless two Weyl points with opposite chiralities meet and annihilate each other.

\section{Discussion}
\label{discussion}
We discuss connections between our work and previous studies.
Reference \onlinecite{Brydon2014} reached a general conclusion that pure electron-phonon interaction in a system with time-reversal and inversion symmetries can generate odd-parity superconductivity, but its instability temperature can not be larger than that of the $s$-wave superconductivity. Our results are consistent with this general statement, and we also show that local Coulomb repulsion can tip the balance in favor of odd-parity pairing. In Ref.~\onlinecite{Wan2014}, Wan and Savrasov presented a first principle study of phonon mediated superconductivity in Cu doped Bi$_2$Se$_3$. Encouragingly, they found that pure electron-phonon interaction does generate odd-parity pairings in both $E_u$ and $A_{2u}$ channels besides the usual even-parity channel. Their calculation indicated that the phonon-mediated instability is stronger in $A_{2u}$ channel compared to $E_u$ channel. In Appendix \ref{appA}, we show that an on-site repulsive interaction in Bi$_2$Se$_3$ generates repulsion in both the $s$-wave channel and  $A_{2u}$ channel, but not in $E_u$ channel. 
In general a finite-range repulsive interaction could also suppress $E_u$ pairing\cite{Fu_Berg}. However, the on-site interaction presumably leads to the most dominant repulsion, which could make $E_u$ pairing more favorable compared to $s$-wave and $A_{2u}$ pairings. It is interesting to reexamine electron-phonon interaction in metal doped Bi$_2$Se$_3$ using {\em ab initio} calculation. In particular, parameters $\gamma_i$, which determine whether nematic or chiral superconductivity is realized in our theory, could be extracted from such a study. In our work, we do not attempt to determine the critical temperature of $E_u$ superconductivity. Such a task requires a detailed knowledge about electron-phonon interaction, which we leave for {\em ab initio} calculation.  The study of Wan and Savrasov\cite{Wan2014} has shown that the electron-phonon interaction is capable of producing a critical temperature of $3\sim 5$ K in the $A_{2u}$ channel.

In summary, we studied odd-parity fluctuations as a possible mechanism for the nematic superconductivity observed in doped Bi$_2$Se$_3$.

\section{Acknowledgment}
F.W. thanks J.W.F. Venderbos for stimulating discussion.
We acknowledge support
from Department of Energy, Office of Basic Energy
Science, Materials Science and Engineering Division.

\appendix
\section{Onsite repulsion in $\text{Bi}_2\text{Se}_3$}
\label{appA}
In this appendix, we show that an on-site repulsive interaction in Bi$_2$Se$_3$ generates repulsive interaction in $s$-wave and $A_{2u}$ pairing channels.
We start from a two-orbital $k\cdot p$ model of Bi$_2$Se$_3$:
\begin{equation}
\mathcal{H}_0(\kk) = M\sigma_x+ v(k_x \tilde{s}_y-k_y \tilde{s}_x)\sigma_z+v_z k_z \sigma_y-\tilde{\mu},
\end{equation}
where $\sigma_a$ and $\tilde{s}_a$ are Pauli matrices respectively in the orbital and spin spaces.
Here $\tilde{\mu}$ and the chemical potential $\mu$ in (\ref{HH0}) are related by $\tilde{\mu}=\mu+M$.
$\mathcal{H}_0(\kk)$ is expressed in the basis $d_{\kk}=(d_{\kk, 1 +},d_{\kk, 1 -},d_{\kk, 2 +},d_{\kk, 2 -})^{\text{T}}$, where the subscript 1 and 2 label the two orbitals, and $\pm$ are the spin indices.  Here the two orbitals are mainly derived from Se $p_z$ orbitals localized on top and bottom layers of the Bi$_2$Se$_3$ unit cell\cite{Zhang2009}. The two orbitals are interchanged under inversion operation. $\mathcal{H}_0(\kk)$ has four bands, corresponding to the two-fold generate valence bands and another two-fold degenerate conduction bands near the band gap.

We consider an on-site repulsive interaction within each orbital:
\begin{equation}
H_U=\frac{2 \tilde{U} }{\Omega}\sum_{\pp \kk \kk'}\sum_{\sigma=1,2} d_{\pp+\kk,\sigma +}^\dagger d_{\pp-\kk,\sigma -}^\dagger d_{\pp-\kk',\sigma -}  d_{\pp+\kk',\sigma +},
\end{equation}
Here $\tilde{U}$ is positive for repulsive interaction.
We decompose $H_U$ into BCS channels:
\begin{equation}
\begin{aligned}
H_U \approx \frac{\tilde{U}}{\Omega} \sum_{\kk,\kk'} \Big\{[\sum_{\sigma} d_{\kk,\sigma \uparrow}^\dagger d_{-\kk,\sigma \downarrow}^\dagger ][\sum_{\sigma'} d_{-\kk',\sigma' \downarrow} d_{\kk',\sigma' \uparrow}]\\
+[\sum_{\sigma} \sigma_z^{(\sigma\sigma)}d_{\kk,\sigma \uparrow}^\dagger d_{-\kk,\sigma \downarrow}^\dagger ][\sum_{\sigma'} \sigma_z^{(\sigma'\sigma')}d_{-\kk',\sigma' \downarrow} d_{\kk',\sigma' \uparrow}]\Big\},
\end{aligned}
\label{HU}
\end{equation}
where the first and second line respectively represent even and odd parity pairing channels.
Finally we project them to the conduction bands\cite{Venderbos2016_pairing}
\begin{equation}
\begin{aligned}
&\sum_{\kk,\sigma} d_{\kk,1 \uparrow}^\dagger d_{-\kk,1 \downarrow}^\dagger + d_{\kk,2 \uparrow}^\dagger d_{-\kk,2 \downarrow}^\dagger\\
\approx & \frac{1}{2}\sum_{\kk}\sum_{\alpha \beta} c_{\kk \alpha}^\dagger \epsilon_{\alpha \beta} c_{-\kk \beta}^\dagger,\\
&\sum_{\kk,\sigma}  d_{\kk,1 \uparrow}^\dagger d_{-\kk,1 \downarrow}^\dagger -d_{\kk,2 \uparrow}^\dagger d_{-\kk,2 \downarrow}^\dagger \\
\approx &\frac{1}{2}\sum_{\kk}\sum_{\alpha \beta} c_{\kk \alpha}^\dagger [\frac{v}{\tilde{\mu}}(k_x s_y-k_y s_x)\epsilon]_{\alpha \beta} c_{-\kk \beta}^\dagger.
\end{aligned}
\label{A2upairing}
\end{equation}
By looking up Table~\ref{tab:form}, it is clear that the odd-parity pairing in (\ref{A2upairing}) belongs to $A_{2u}$ representation.

\section{Odd-parity fluctuation in $A_{1u}$ representation}
\label{appB}
In Bi$_2$Se$_3$, there is no Brillouin-zone-center phonon mode in $A_{1u}$ representation\cite{Wang2012Phonon}.  Nevertheless, we can still theoretically study superconductivity induced by odd-parity particle-hole fluctuation in $A_{1u}$ representation.
The procedure is parallel to that presented in Sec.~\ref{FlucSuc}.
The main difference is the form factor:
\begin{equation}
\Gamma_1(\kk)=\gamma_1 \Gamma_1^{(1)}(\kk)+\gamma_2 \Gamma_1^{(2)}(\kk),
\label{Gamma1}
\end{equation}
where $\Gamma_1^{(1)}(\kk)$ and $\Gamma_1^{(2)}(\kk)$, given in Table~\ref{tab:form}, are two basis functions in $A_{1u}$ representation up to first order in $\kk$.

The effective interaction induced by $A_{1u}$ fluctuation can again be decomposed into even and odd parity pairing channels:
\begin{equation}
\begin{aligned}
H_e=&\frac{V_0}{\Omega}\sum_{\kk,\kk'}[g_0(\kk)+g_0(\kk')][\frac{1}{2}\epsilon_{\alpha\beta}c^\dagger_{\kk\alpha}c^\dagger_{-\kk\beta}][\frac{1}{2}\epsilon^{\dagger}_{\gamma \delta}c_{-\kk'\gamma}   c_{\kk'\delta}],\\
H_{o}=&\frac{V_0}{\Omega}\Big\{(\gamma_1 \hat{F}_{1}^{(1)}+\gamma_2 \hat{F}_{1}^{(2)} )^\dagger(\gamma_1 \hat{F}_{1}^{(1)}+\gamma_2 \hat{F}_{1}^{(2)})\\
-&\gamma_1^2 \hat{F}_{2}^{(1)\dagger}\hat{F}_{2}^{(1)}\\
-&\sum_{a=x,y}(\frac{\gamma_1}{\sqrt{2}}\hat{F}_a^{(1)}-\gamma_2\hat{F}_a^{(2)})^{\dagger}(\frac{\gamma_1}{\sqrt{2}}\hat{F}_a^{(1)}-\gamma_2\hat{F}_a^{(2)})
\Big\},
\end{aligned}
\label{Heo}
\end{equation}
where $H_e$ describes attractive interaction in even-parity channel, and the form factor is $g_0(\kk)=\gamma_1^2(\hat{k}_x^2+\hat{k}_y^2)/4+\gamma_2^2 \hat{k}_z^2$/2, which does not include repulsive interaction in the $s$-wave channel.
In $H_o$ of Eq. (\ref{Heo}), $A_{1u}$ pairing channel has attractive interaction, while the other two odd-parity channels are repulsive.  

The critical temperature in the even-parity and odd-parity $A_{1u}$ channels are separately given by the corresponding linearized gap equations:
\begin{equation}
\begin{aligned}
&|V_0|\chi_s(T_{c,s})=1,\quad\quad |V_0|\chi_p(T_{c,p})=1,\\
&\frac{\chi_s(T)}{\chi_0(T)}=\frac{\gamma_1^2+\gamma_2^2}{6}+\sqrt{\frac{1}{60} (2\gamma_1^4+2\gamma_1^2\gamma_2^2+3\gamma_2^4)},\\
&\frac{\chi_p(T)}{\chi_0(T)}= \frac{\gamma_1^2+\gamma_2^2}{3}.
\end{aligned}
\end{equation}
The ratio $\chi_p/\chi_s$ takes its minimum value 0.85 when $\gamma_1=0$, and its maximum value 1 when $\gamma_1/\gamma_2=\sqrt{2}$.
Therefore, $s$-wave and $A_{1u}$ pairings can have the same critical temperature even without considering the repulsive interaction in the $s$-wave channel\cite{Kozii2015, Wang2016, Wang_Fu}. 

\section{Odd-parity fluctuation in $A_{2u}$ representation}
\label{appC}
There are $A_{2u}$ phonon modes at the Brillouin zone center in Bi$_2$Se$_3$. The corresponding form factor has only one basis function to linear order in $\kk$:
\begin{equation}
\Gamma_2(\kk) = \gamma_1 \Gamma_2^{(1)}(\kk)= \frac{\gamma_1}{\sqrt{2}}(\hatk_x s_y - \hatk_y s_x).
\end{equation}

In the effective interaction, the even-parity part $H_e$ takes similar form as that in (\ref{Heo}), but the form factor $g_o(\kk)$ is replaced  by 
$\gamma_1^2(\hat{k}_x^2+\hat{k}_y^2)/4$. The odd-parity part $H_o$ is given by:
\begin{equation}
\begin{aligned}
H_{o}=\frac{\gamma_1^2 V_0}{\Omega}\Big\{&- \hat{F}_{1}^{(1)\dagger} \hat{F}_{1}^{(1)}
+ \hat{F}_{2}^{(1)\dagger} \hat{F}_{2}^{(1)}\\
&- \frac{1}{2} \sum_{a=x,y} \hat{F}_a^{(1)\dagger} \hat{F}_a^{(1)} 
\Big\},
\end{aligned}
\end{equation}
where only the $A_{2u}$ pairing channel has an attractive interaction.

The linearized gap equations for even-parity and $A_{2u}$ channels are respectively expressed as:
\begin{equation}
\begin{aligned}
&\gamma_1^2|V_0|\chi_s(T_{c,s})=1,\quad\quad\gamma_1^2|V_0|\chi_p(T_{c,p})=1,\\
&\frac{\chi_s(T)}{\chi_0(T)}= \frac{1}{6}+\sqrt{\frac{1}{30}},  \quad \frac{\chi_p(T)}{\chi_0(T)}= \frac{1}{3}.
\end{aligned}
\label{A2ugap}
\end{equation}
Here the ratio $\chi_p/\chi_s$ is about 0.95, indicating that the critical temperature for the two channels can be comparable. 
For simplicity, the gap equations in (\ref{A2ugap}) do not include the repulsive interaction discussed in Appendix~\ref{appA}. 

\bibliographystyle{apsrev4-1}
\bibliography{refs}

\begin{thebibliography}{51}%
\makeatletter
\providecommand \@ifxundefined [1]{%
 \@ifx{#1\undefined}
}%
\providecommand \@ifnum [1]{%
 \ifnum #1\expandafter \@firstoftwo
 \else \expandafter \@secondoftwo
 \fi
}%
\providecommand \@ifx [1]{%
 \ifx #1\expandafter \@firstoftwo
 \else \expandafter \@secondoftwo
 \fi
}%
\providecommand \natexlab [1]{#1}%
\providecommand \enquote  [1]{``#1''}%
\providecommand \bibnamefont  [1]{#1}%
\providecommand \bibfnamefont [1]{#1}%
\providecommand \citenamefont [1]{#1}%
\providecommand \href@noop [0]{\@secondoftwo}%
\providecommand \href [0]{\begingroup \@sanitize@url \@href}%
\providecommand \@href[1]{\@@startlink{#1}\@@href}%
\providecommand \@@href[1]{\endgroup#1\@@endlink}%
\providecommand \@sanitize@url [0]{\catcode `\\12\catcode `\$12\catcode
  `\&12\catcode `\#12\catcode `\^12\catcode `\_12\catcode `\%12\relax}%
\providecommand \@@startlink[1]{}%
\providecommand \@@endlink[0]{}%
\providecommand \url  [0]{\begingroup\@sanitize@url \@url }%
\providecommand \@url [1]{\endgroup\@href {#1}{\urlprefix }}%
\providecommand \urlprefix  [0]{URL }%
\providecommand \Eprint [0]{\href }%
\providecommand \doibase [0]{http://dx.doi.org/}%
\providecommand \selectlanguage [0]{\@gobble}%
\providecommand \bibinfo  [0]{\@secondoftwo}%
\providecommand \bibfield  [0]{\@secondoftwo}%
\providecommand \translation [1]{[#1]}%
\providecommand \BibitemOpen [0]{}%
\providecommand \bibitemStop [0]{}%
\providecommand \bibitemNoStop [0]{.\EOS\space}%
\providecommand \EOS [0]{\spacefactor3000\relax}%
\providecommand \BibitemShut  [1]{\csname bibitem#1\endcsname}%
\let\auto@bib@innerbib\@empty
\bibitem [{\citenamefont {Kane}\ and\ \citenamefont {Mele}(2005)}]{KaneMele}%
  \BibitemOpen
  \bibfield  {author} {\bibinfo {author} {\bibfnamefont {C.~L.}\ \bibnamefont
  {Kane}}\ and\ \bibinfo {author} {\bibfnamefont {E.~J.}\ \bibnamefont
  {Mele}},\ }\href {\doibase 10.1103/PhysRevLett.95.146802} {\bibfield
  {journal} {\bibinfo  {journal} {Phys. Rev. Lett.}\ }\textbf {\bibinfo
  {volume} {95}},\ \bibinfo {pages} {146802} (\bibinfo {year}
  {2005})}\BibitemShut {NoStop}%
\bibitem [{\citenamefont {Fu}\ \emph {et~al.}(2007)\citenamefont {Fu},
  \citenamefont {Kane},\ and\ \citenamefont {Mele}}]{FuKaneMele}%
  \BibitemOpen
  \bibfield  {author} {\bibinfo {author} {\bibfnamefont {L.}~\bibnamefont
  {Fu}}, \bibinfo {author} {\bibfnamefont {C.~L.}\ \bibnamefont {Kane}}, \ and\
  \bibinfo {author} {\bibfnamefont {E.~J.}\ \bibnamefont {Mele}},\ }\href
  {\doibase 10.1103/PhysRevLett.98.106803} {\bibfield  {journal} {\bibinfo
  {journal} {Phys. Rev. Lett.}\ }\textbf {\bibinfo {volume} {98}},\ \bibinfo
  {pages} {106803} (\bibinfo {year} {2007})}\BibitemShut {NoStop}%
\bibitem [{\citenamefont {Hasan}\ and\ \citenamefont {Kane}(2010)}]{HasanKane}%
  \BibitemOpen
  \bibfield  {author} {\bibinfo {author} {\bibfnamefont {M.~Z.}\ \bibnamefont
  {Hasan}}\ and\ \bibinfo {author} {\bibfnamefont {C.~L.}\ \bibnamefont
  {Kane}},\ }\href {\doibase 10.1103/RevModPhys.82.3045} {\bibfield  {journal}
  {\bibinfo  {journal} {Rev. Mod. Phys.}\ }\textbf {\bibinfo {volume} {82}},\
  \bibinfo {pages} {3045} (\bibinfo {year} {2010})}\BibitemShut {NoStop}%
\bibitem [{\citenamefont {Qi}\ and\ \citenamefont {Zhang}(2011)}]{QiZhang}%
  \BibitemOpen
  \bibfield  {author} {\bibinfo {author} {\bibfnamefont {X.-L.}\ \bibnamefont
  {Qi}}\ and\ \bibinfo {author} {\bibfnamefont {S.-C.}\ \bibnamefont {Zhang}},\
  }\href {\doibase 10.1103/RevModPhys.83.1057} {\bibfield  {journal} {\bibinfo
  {journal} {Rev. Mod. Phys.}\ }\textbf {\bibinfo {volume} {83}},\ \bibinfo
  {pages} {1057} (\bibinfo {year} {2011})}\BibitemShut {NoStop}%
\bibitem [{\citenamefont {Liu}\ \emph {et~al.}(2014)\citenamefont {Liu},
  \citenamefont {Jiang}, \citenamefont {Zhou}, \citenamefont {Wang},
  \citenamefont {Zhang}, \citenamefont {Weng}, \citenamefont {Prabhakaran},
  \citenamefont {Mo}, \citenamefont {Peng}, \citenamefont {Dudin} \emph
  {et~al.}}]{Liu2014}%
  \BibitemOpen
  \bibfield  {author} {\bibinfo {author} {\bibfnamefont {Z.}~\bibnamefont
  {Liu}}, \bibinfo {author} {\bibfnamefont {J.}~\bibnamefont {Jiang}}, \bibinfo
  {author} {\bibfnamefont {B.}~\bibnamefont {Zhou}}, \bibinfo {author}
  {\bibfnamefont {Z.}~\bibnamefont {Wang}}, \bibinfo {author} {\bibfnamefont
  {Y.}~\bibnamefont {Zhang}}, \bibinfo {author} {\bibfnamefont
  {H.}~\bibnamefont {Weng}}, \bibinfo {author} {\bibfnamefont {D.}~\bibnamefont
  {Prabhakaran}}, \bibinfo {author} {\bibfnamefont {S.}~\bibnamefont {Mo}},
  \bibinfo {author} {\bibfnamefont {H.}~\bibnamefont {Peng}}, \bibinfo {author}
  {\bibfnamefont {P.}~\bibnamefont {Dudin}},  \emph {et~al.},\ }\href
  {https://www.nature.com/nmat/journal/v13/n7/abs/nmat3990.html} {\bibfield
  {journal} {\bibinfo  {journal} {Nat. Mater.}\ }\textbf {\bibinfo {volume}
  {13}},\ \bibinfo {pages} {677} (\bibinfo {year} {2014})}\BibitemShut
  {NoStop}%
\bibitem [{\citenamefont {Xu}\ \emph {et~al.}(2015)\citenamefont {Xu},
  \citenamefont {Belopolski}, \citenamefont {Alidoust}, \citenamefont
  {Neupane}, \citenamefont {Bian}, \citenamefont {Zhang}, \citenamefont
  {Sankar}, \citenamefont {Chang}, \citenamefont {Yuan}, \citenamefont {Lee}
  \emph {et~al.}}]{Xu2015}%
  \BibitemOpen
  \bibfield  {author} {\bibinfo {author} {\bibfnamefont {S.-Y.}\ \bibnamefont
  {Xu}}, \bibinfo {author} {\bibfnamefont {I.}~\bibnamefont {Belopolski}},
  \bibinfo {author} {\bibfnamefont {N.}~\bibnamefont {Alidoust}}, \bibinfo
  {author} {\bibfnamefont {M.}~\bibnamefont {Neupane}}, \bibinfo {author}
  {\bibfnamefont {G.}~\bibnamefont {Bian}}, \bibinfo {author} {\bibfnamefont
  {C.}~\bibnamefont {Zhang}}, \bibinfo {author} {\bibfnamefont
  {R.}~\bibnamefont {Sankar}}, \bibinfo {author} {\bibfnamefont
  {G.}~\bibnamefont {Chang}}, \bibinfo {author} {\bibfnamefont
  {Z.}~\bibnamefont {Yuan}}, \bibinfo {author} {\bibfnamefont {C.-C.}\
  \bibnamefont {Lee}},  \emph {et~al.},\ }\href
  {http://science.sciencemag.org/content/349/6248/613} {\bibfield  {journal}
  {\bibinfo  {journal} {Science}\ }\textbf {\bibinfo {volume} {349}},\ \bibinfo
  {pages} {613} (\bibinfo {year} {2015})}\BibitemShut {NoStop}%
\bibitem [{\citenamefont {Fu}\ and\ \citenamefont {Kane}(2008)}]{Fu_Kane}%
  \BibitemOpen
  \bibfield  {author} {\bibinfo {author} {\bibfnamefont {L.}~\bibnamefont
  {Fu}}\ and\ \bibinfo {author} {\bibfnamefont {C.~L.}\ \bibnamefont {Kane}},\
  }\href {\doibase 10.1103/PhysRevLett.100.096407} {\bibfield  {journal}
  {\bibinfo  {journal} {Phys. Rev. Lett.}\ }\textbf {\bibinfo {volume} {100}},\
  \bibinfo {pages} {096407} (\bibinfo {year} {2008})}\BibitemShut {NoStop}%
\bibitem [{\citenamefont {Qi}\ \emph {et~al.}(2009)\citenamefont {Qi},
  \citenamefont {Hughes}, \citenamefont {Raghu},\ and\ \citenamefont
  {Zhang}}]{Qi2009}%
  \BibitemOpen
  \bibfield  {author} {\bibinfo {author} {\bibfnamefont {X.-L.}\ \bibnamefont
  {Qi}}, \bibinfo {author} {\bibfnamefont {T.~L.}\ \bibnamefont {Hughes}},
  \bibinfo {author} {\bibfnamefont {S.}~\bibnamefont {Raghu}}, \ and\ \bibinfo
  {author} {\bibfnamefont {S.-C.}\ \bibnamefont {Zhang}},\ }\href {\doibase
  10.1103/PhysRevLett.102.187001} {\bibfield  {journal} {\bibinfo  {journal}
  {Phys. Rev. Lett.}\ }\textbf {\bibinfo {volume} {102}},\ \bibinfo {pages}
  {187001} (\bibinfo {year} {2009})}\BibitemShut {NoStop}%
\bibitem [{\citenamefont {Sau}\ \emph {et~al.}(2010)\citenamefont {Sau},
  \citenamefont {Lutchyn}, \citenamefont {Tewari},\ and\ \citenamefont
  {Das~Sarma}}]{Sau2010}%
  \BibitemOpen
  \bibfield  {author} {\bibinfo {author} {\bibfnamefont {J.~D.}\ \bibnamefont
  {Sau}}, \bibinfo {author} {\bibfnamefont {R.~M.}\ \bibnamefont {Lutchyn}},
  \bibinfo {author} {\bibfnamefont {S.}~\bibnamefont {Tewari}}, \ and\ \bibinfo
  {author} {\bibfnamefont {S.}~\bibnamefont {Das~Sarma}},\ }\href {\doibase
  10.1103/PhysRevLett.104.040502} {\bibfield  {journal} {\bibinfo  {journal}
  {Phys. Rev. Lett.}\ }\textbf {\bibinfo {volume} {104}},\ \bibinfo {pages}
  {040502} (\bibinfo {year} {2010})}\BibitemShut {NoStop}%
\bibitem [{\citenamefont {Lutchyn}\ \emph {et~al.}(2010)\citenamefont
  {Lutchyn}, \citenamefont {Sau},\ and\ \citenamefont
  {Das~Sarma}}]{Lutchyn2010}%
  \BibitemOpen
  \bibfield  {author} {\bibinfo {author} {\bibfnamefont {R.~M.}\ \bibnamefont
  {Lutchyn}}, \bibinfo {author} {\bibfnamefont {J.~D.}\ \bibnamefont {Sau}}, \
  and\ \bibinfo {author} {\bibfnamefont {S.}~\bibnamefont {Das~Sarma}},\ }\href
  {\doibase 10.1103/PhysRevLett.105.077001} {\bibfield  {journal} {\bibinfo
  {journal} {Phys. Rev. Lett.}\ }\textbf {\bibinfo {volume} {105}},\ \bibinfo
  {pages} {077001} (\bibinfo {year} {2010})}\BibitemShut {NoStop}%
\bibitem [{\citenamefont {Nadj-Perge}\ \emph {et~al.}(2014)\citenamefont
  {Nadj-Perge}, \citenamefont {Drozdov}, \citenamefont {Li}, \citenamefont
  {Chen}, \citenamefont {Jeon}, \citenamefont {Seo}, \citenamefont {MacDonald},
  \citenamefont {Bernevig},\ and\ \citenamefont {Yazdani}}]{nadj2014}%
  \BibitemOpen
  \bibfield  {author} {\bibinfo {author} {\bibfnamefont {S.}~\bibnamefont
  {Nadj-Perge}}, \bibinfo {author} {\bibfnamefont {I.~K.}\ \bibnamefont
  {Drozdov}}, \bibinfo {author} {\bibfnamefont {J.}~\bibnamefont {Li}},
  \bibinfo {author} {\bibfnamefont {H.}~\bibnamefont {Chen}}, \bibinfo {author}
  {\bibfnamefont {S.}~\bibnamefont {Jeon}}, \bibinfo {author} {\bibfnamefont
  {J.}~\bibnamefont {Seo}}, \bibinfo {author} {\bibfnamefont {A.~H.}\
  \bibnamefont {MacDonald}}, \bibinfo {author} {\bibfnamefont {B.~A.}\
  \bibnamefont {Bernevig}}, \ and\ \bibinfo {author} {\bibfnamefont
  {A.}~\bibnamefont {Yazdani}},\ }\href@noop {} {\bibfield  {journal} {\bibinfo
   {journal} {Science}\ }\textbf {\bibinfo {volume} {346}},\ \bibinfo {pages}
  {602} (\bibinfo {year} {2014})}\BibitemShut {NoStop}%
\bibitem [{\citenamefont {Albrecht}\ \emph {et~al.}(2016)\citenamefont
  {Albrecht}, \citenamefont {Higginbotham}, \citenamefont {Madsen},
  \citenamefont {Kuemmeth}, \citenamefont {Jespersen}, \citenamefont
  {Nyg{\aa}rd}, \citenamefont {Krogstrup},\ and\ \citenamefont
  {Marcus}}]{albrecht2016}%
  \BibitemOpen
  \bibfield  {author} {\bibinfo {author} {\bibfnamefont {S.~M.}\ \bibnamefont
  {Albrecht}}, \bibinfo {author} {\bibfnamefont {A.}~\bibnamefont
  {Higginbotham}}, \bibinfo {author} {\bibfnamefont {M.}~\bibnamefont
  {Madsen}}, \bibinfo {author} {\bibfnamefont {F.}~\bibnamefont {Kuemmeth}},
  \bibinfo {author} {\bibfnamefont {T.~S.}\ \bibnamefont {Jespersen}}, \bibinfo
  {author} {\bibfnamefont {J.}~\bibnamefont {Nyg{\aa}rd}}, \bibinfo {author}
  {\bibfnamefont {P.}~\bibnamefont {Krogstrup}}, \ and\ \bibinfo {author}
  {\bibfnamefont {C.}~\bibnamefont {Marcus}},\ }\href
  {http://www.nature.com/nature/journal/v531/n7593/full/nature17162.html}
  {\bibfield  {journal} {\bibinfo  {journal} {Nature}\ }\textbf {\bibinfo
  {volume} {531}},\ \bibinfo {pages} {206} (\bibinfo {year}
  {2016})}\BibitemShut {NoStop}%
\bibitem [{\citenamefont {Matano}\ \emph {et~al.}(2016)\citenamefont {Matano},
  \citenamefont {Kriener}, \citenamefont {Segawa}, \citenamefont {Ando},\ and\
  \citenamefont {Zheng}}]{matano2016spin}%
  \BibitemOpen
  \bibfield  {author} {\bibinfo {author} {\bibfnamefont {K.}~\bibnamefont
  {Matano}}, \bibinfo {author} {\bibfnamefont {M.}~\bibnamefont {Kriener}},
  \bibinfo {author} {\bibfnamefont {K.}~\bibnamefont {Segawa}}, \bibinfo
  {author} {\bibfnamefont {Y.}~\bibnamefont {Ando}}, \ and\ \bibinfo {author}
  {\bibfnamefont {G.-q.}\ \bibnamefont {Zheng}},\ }\href
  {http://www.nature.com/nphys/journal/v12/n9/abs/nphys3781.html} {\bibfield
  {journal} {\bibinfo  {journal} {Nat. Phys.}\ }\textbf {\bibinfo {volume}
  {12}},\ \bibinfo {pages} {852} (\bibinfo {year} {2016})}\BibitemShut
  {NoStop}%
\bibitem [{\citenamefont {Yonezawa}\ \emph {et~al.}(2017)\citenamefont
  {Yonezawa}, \citenamefont {Tajiri}, \citenamefont {Nakata}, \citenamefont
  {Nagai}, \citenamefont {Wang}, \citenamefont {Segawa}, \citenamefont {Ando},\
  and\ \citenamefont {Maeno}}]{yonezawa2017}%
  \BibitemOpen
  \bibfield  {author} {\bibinfo {author} {\bibfnamefont {S.}~\bibnamefont
  {Yonezawa}}, \bibinfo {author} {\bibfnamefont {K.}~\bibnamefont {Tajiri}},
  \bibinfo {author} {\bibfnamefont {S.}~\bibnamefont {Nakata}}, \bibinfo
  {author} {\bibfnamefont {Y.}~\bibnamefont {Nagai}}, \bibinfo {author}
  {\bibfnamefont {Z.}~\bibnamefont {Wang}}, \bibinfo {author} {\bibfnamefont
  {K.}~\bibnamefont {Segawa}}, \bibinfo {author} {\bibfnamefont
  {Y.}~\bibnamefont {Ando}}, \ and\ \bibinfo {author} {\bibfnamefont
  {Y.}~\bibnamefont {Maeno}},\ }\href
  {http://www.nature.com/nphys/journal/v13/n2/full/nphys3907.html#ref6}
  {\bibfield  {journal} {\bibinfo  {journal} {Nat. Phys.}\ }\textbf {\bibinfo
  {volume} {13}},\ \bibinfo {pages} {123} (\bibinfo {year} {2017})}\BibitemShut
  {NoStop}%
\bibitem [{\citenamefont {Pan}\ \emph {et~al.}(2016)\citenamefont {Pan},
  \citenamefont {Nikitin}, \citenamefont {Araizi}, \citenamefont {Huang},
  \citenamefont {Matsushita}, \citenamefont {Naka},\ and\ \citenamefont
  {De~Visser}}]{pan2016rotational}%
  \BibitemOpen
  \bibfield  {author} {\bibinfo {author} {\bibfnamefont {Y.}~\bibnamefont
  {Pan}}, \bibinfo {author} {\bibfnamefont {A.}~\bibnamefont {Nikitin}},
  \bibinfo {author} {\bibfnamefont {G.}~\bibnamefont {Araizi}}, \bibinfo
  {author} {\bibfnamefont {Y.}~\bibnamefont {Huang}}, \bibinfo {author}
  {\bibfnamefont {Y.}~\bibnamefont {Matsushita}}, \bibinfo {author}
  {\bibfnamefont {T.}~\bibnamefont {Naka}}, \ and\ \bibinfo {author}
  {\bibfnamefont {A.}~\bibnamefont {De~Visser}},\ }\href
  {http://www.nature.com/articles/srep28632} {\bibfield  {journal} {\bibinfo
  {journal} {Sci. Rep.}\ }\textbf {\bibinfo {volume} {6}},\ \bibinfo {pages}
  {28632} (\bibinfo {year} {2016})}\BibitemShut {NoStop}%
\bibitem [{\citenamefont {Asaba}\ \emph {et~al.}(2017)\citenamefont {Asaba},
  \citenamefont {Lawson}, \citenamefont {Tinsman}, \citenamefont {Chen},
  \citenamefont {Corbae}, \citenamefont {Li}, \citenamefont {Qiu},
  \citenamefont {Hor}, \citenamefont {Fu},\ and\ \citenamefont
  {Li}}]{Asaba2017}%
  \BibitemOpen
  \bibfield  {author} {\bibinfo {author} {\bibfnamefont {T.}~\bibnamefont
  {Asaba}}, \bibinfo {author} {\bibfnamefont {B.~J.}\ \bibnamefont {Lawson}},
  \bibinfo {author} {\bibfnamefont {C.}~\bibnamefont {Tinsman}}, \bibinfo
  {author} {\bibfnamefont {L.}~\bibnamefont {Chen}}, \bibinfo {author}
  {\bibfnamefont {P.}~\bibnamefont {Corbae}}, \bibinfo {author} {\bibfnamefont
  {G.}~\bibnamefont {Li}}, \bibinfo {author} {\bibfnamefont {Y.}~\bibnamefont
  {Qiu}}, \bibinfo {author} {\bibfnamefont {Y.~S.}\ \bibnamefont {Hor}},
  \bibinfo {author} {\bibfnamefont {L.}~\bibnamefont {Fu}}, \ and\ \bibinfo
  {author} {\bibfnamefont {L.}~\bibnamefont {Li}},\ }\href {\doibase
  10.1103/PhysRevX.7.011009} {\bibfield  {journal} {\bibinfo  {journal} {Phys.
  Rev. X}\ }\textbf {\bibinfo {volume} {7}},\ \bibinfo {pages} {011009}
  (\bibinfo {year} {2017})}\BibitemShut {NoStop}%
\bibitem [{\citenamefont {Kriener}\ \emph {et~al.}(2011)\citenamefont
  {Kriener}, \citenamefont {Segawa}, \citenamefont {Ren}, \citenamefont
  {Sasaki},\ and\ \citenamefont {Ando}}]{Ando2011}%
  \BibitemOpen
  \bibfield  {author} {\bibinfo {author} {\bibfnamefont {M.}~\bibnamefont
  {Kriener}}, \bibinfo {author} {\bibfnamefont {K.}~\bibnamefont {Segawa}},
  \bibinfo {author} {\bibfnamefont {Z.}~\bibnamefont {Ren}}, \bibinfo {author}
  {\bibfnamefont {S.}~\bibnamefont {Sasaki}}, \ and\ \bibinfo {author}
  {\bibfnamefont {Y.}~\bibnamefont {Ando}},\ }\href {\doibase
  10.1103/PhysRevLett.106.127004} {\bibfield  {journal} {\bibinfo  {journal}
  {Phys. Rev. Lett.}\ }\textbf {\bibinfo {volume} {106}},\ \bibinfo {pages}
  {127004} (\bibinfo {year} {2011})}\BibitemShut {NoStop}%
\bibitem [{\citenamefont {Smylie}\ \emph {et~al.}(2016)\citenamefont {Smylie},
  \citenamefont {Claus}, \citenamefont {Welp}, \citenamefont {Kwok},
  \citenamefont {Qiu}, \citenamefont {Hor},\ and\ \citenamefont
  {Snezhko}}]{Smylie2016}%
  \BibitemOpen
  \bibfield  {author} {\bibinfo {author} {\bibfnamefont {M.~P.}\ \bibnamefont
  {Smylie}}, \bibinfo {author} {\bibfnamefont {H.}~\bibnamefont {Claus}},
  \bibinfo {author} {\bibfnamefont {U.}~\bibnamefont {Welp}}, \bibinfo {author}
  {\bibfnamefont {W.-K.}\ \bibnamefont {Kwok}}, \bibinfo {author}
  {\bibfnamefont {Y.}~\bibnamefont {Qiu}}, \bibinfo {author} {\bibfnamefont
  {Y.~S.}\ \bibnamefont {Hor}}, \ and\ \bibinfo {author} {\bibfnamefont
  {A.}~\bibnamefont {Snezhko}},\ }\href {\doibase 10.1103/PhysRevB.94.180510}
  {\bibfield  {journal} {\bibinfo  {journal} {Phys. Rev. B}\ }\textbf {\bibinfo
  {volume} {94}},\ \bibinfo {pages} {180510} (\bibinfo {year}
  {2016})}\BibitemShut {NoStop}%
\bibitem [{\citenamefont {Fu}(2014)}]{Fu_Nematic}%
  \BibitemOpen
  \bibfield  {author} {\bibinfo {author} {\bibfnamefont {L.}~\bibnamefont
  {Fu}},\ }\href {\doibase 10.1103/PhysRevB.90.100509} {\bibfield  {journal}
  {\bibinfo  {journal} {Phys. Rev. B}\ }\textbf {\bibinfo {volume} {90}},\
  \bibinfo {pages} {100509} (\bibinfo {year} {2014})}\BibitemShut {NoStop}%
\bibitem [{\citenamefont {Sasaki}\ \emph {et~al.}(2011)\citenamefont {Sasaki},
  \citenamefont {Kriener}, \citenamefont {Segawa}, \citenamefont {Yada},
  \citenamefont {Tanaka}, \citenamefont {Sato},\ and\ \citenamefont
  {Ando}}]{Sasaki2011}%
  \BibitemOpen
  \bibfield  {author} {\bibinfo {author} {\bibfnamefont {S.}~\bibnamefont
  {Sasaki}}, \bibinfo {author} {\bibfnamefont {M.}~\bibnamefont {Kriener}},
  \bibinfo {author} {\bibfnamefont {K.}~\bibnamefont {Segawa}}, \bibinfo
  {author} {\bibfnamefont {K.}~\bibnamefont {Yada}}, \bibinfo {author}
  {\bibfnamefont {Y.}~\bibnamefont {Tanaka}}, \bibinfo {author} {\bibfnamefont
  {M.}~\bibnamefont {Sato}}, \ and\ \bibinfo {author} {\bibfnamefont
  {Y.}~\bibnamefont {Ando}},\ }\href {\doibase 10.1103/PhysRevLett.107.217001}
  {\bibfield  {journal} {\bibinfo  {journal} {Phys. Rev. Lett.}\ }\textbf
  {\bibinfo {volume} {107}},\ \bibinfo {pages} {217001} (\bibinfo {year}
  {2011})}\BibitemShut {NoStop}%
\bibitem [{\citenamefont {Levy}\ \emph {et~al.}(2013)\citenamefont {Levy},
  \citenamefont {Zhang}, \citenamefont {Ha}, \citenamefont {Sharifi},
  \citenamefont {Talin}, \citenamefont {Kuk},\ and\ \citenamefont
  {Stroscio}}]{Levy2013}%
  \BibitemOpen
  \bibfield  {author} {\bibinfo {author} {\bibfnamefont {N.}~\bibnamefont
  {Levy}}, \bibinfo {author} {\bibfnamefont {T.}~\bibnamefont {Zhang}},
  \bibinfo {author} {\bibfnamefont {J.}~\bibnamefont {Ha}}, \bibinfo {author}
  {\bibfnamefont {F.}~\bibnamefont {Sharifi}}, \bibinfo {author} {\bibfnamefont
  {A.~A.}\ \bibnamefont {Talin}}, \bibinfo {author} {\bibfnamefont
  {Y.}~\bibnamefont {Kuk}}, \ and\ \bibinfo {author} {\bibfnamefont {J.~A.}\
  \bibnamefont {Stroscio}},\ }\href {\doibase 10.1103/PhysRevLett.110.117001}
  {\bibfield  {journal} {\bibinfo  {journal} {Phys. Rev. Lett.}\ }\textbf
  {\bibinfo {volume} {110}},\ \bibinfo {pages} {117001} (\bibinfo {year}
  {2013})}\BibitemShut {NoStop}%
\bibitem [{\citenamefont {Hashimoto}\ \emph {et~al.}(2013)\citenamefont
  {Hashimoto}, \citenamefont {Yada}, \citenamefont {Yamakage}, \citenamefont
  {Sato},\ and\ \citenamefont {Tanaka}}]{hashimoto2013bulk}%
  \BibitemOpen
  \bibfield  {author} {\bibinfo {author} {\bibfnamefont {T.}~\bibnamefont
  {Hashimoto}}, \bibinfo {author} {\bibfnamefont {K.}~\bibnamefont {Yada}},
  \bibinfo {author} {\bibfnamefont {A.}~\bibnamefont {Yamakage}}, \bibinfo
  {author} {\bibfnamefont {M.}~\bibnamefont {Sato}}, \ and\ \bibinfo {author}
  {\bibfnamefont {Y.}~\bibnamefont {Tanaka}},\ }\href
  {http://journals.jps.jp/doi/abs/10.7566/JPSJ.82.044704} {\bibfield  {journal}
  {\bibinfo  {journal} {J. Phys. Soc. Jpn.}\ }\textbf {\bibinfo {volume}
  {82}},\ \bibinfo {pages} {044704} (\bibinfo {year} {2013})}\BibitemShut
  {NoStop}%
\bibitem [{\citenamefont {Nagai}\ and\ \citenamefont {Ota}(2016)}]{Nagai2016}%
  \BibitemOpen
  \bibfield  {author} {\bibinfo {author} {\bibfnamefont {Y.}~\bibnamefont
  {Nagai}}\ and\ \bibinfo {author} {\bibfnamefont {Y.}~\bibnamefont {Ota}},\
  }\href {\doibase 10.1103/PhysRevB.94.134516} {\bibfield  {journal} {\bibinfo
  {journal} {Phys. Rev. B}\ }\textbf {\bibinfo {volume} {94}},\ \bibinfo
  {pages} {134516} (\bibinfo {year} {2016})}\BibitemShut {NoStop}%
\bibitem [{\citenamefont {Venderbos}\ \emph
  {et~al.}(2016{\natexlab{a}})\citenamefont {Venderbos}, \citenamefont
  {Kozii},\ and\ \citenamefont {Fu}}]{Venderbos2016}%
  \BibitemOpen
  \bibfield  {author} {\bibinfo {author} {\bibfnamefont {J.~W.~F.}\
  \bibnamefont {Venderbos}}, \bibinfo {author} {\bibfnamefont {V.}~\bibnamefont
  {Kozii}}, \ and\ \bibinfo {author} {\bibfnamefont {L.}~\bibnamefont {Fu}},\
  }\href {\doibase 10.1103/PhysRevB.94.094522} {\bibfield  {journal} {\bibinfo
  {journal} {Phys. Rev. B}\ }\textbf {\bibinfo {volume} {94}},\ \bibinfo
  {pages} {094522} (\bibinfo {year} {2016}{\natexlab{a}})}\BibitemShut
  {NoStop}%
\bibitem [{\citenamefont {Yang}\ \emph {et~al.}(2014)\citenamefont {Yang},
  \citenamefont {Pan},\ and\ \citenamefont {Zhang}}]{Zhang_2014}%
  \BibitemOpen
  \bibfield  {author} {\bibinfo {author} {\bibfnamefont {S.~A.}\ \bibnamefont
  {Yang}}, \bibinfo {author} {\bibfnamefont {H.}~\bibnamefont {Pan}}, \ and\
  \bibinfo {author} {\bibfnamefont {F.}~\bibnamefont {Zhang}},\ }\href
  {\doibase 10.1103/PhysRevLett.113.046401} {\bibfield  {journal} {\bibinfo
  {journal} {Phys. Rev. Lett.}\ }\textbf {\bibinfo {volume} {113}},\ \bibinfo
  {pages} {046401} (\bibinfo {year} {2014})}\BibitemShut {NoStop}%
\bibitem [{\citenamefont {Wu}\ and\ \citenamefont {Martin}(2017)}]{Wu2017}%
  \BibitemOpen
  \bibfield  {author} {\bibinfo {author} {\bibfnamefont {F.}~\bibnamefont
  {Wu}}\ and\ \bibinfo {author} {\bibfnamefont {I.}~\bibnamefont {Martin}},\
  }\href {\doibase 10.1103/PhysRevB.95.224503} {\bibfield  {journal} {\bibinfo
  {journal} {Phys. Rev. B}\ }\textbf {\bibinfo {volume} {95}},\ \bibinfo
  {pages} {224503} (\bibinfo {year} {2017})}\BibitemShut {NoStop}%
\bibitem [{\citenamefont {Zyuzin}\ \emph {et~al.}()\citenamefont {Zyuzin},
  \citenamefont {Garaud},\ and\ \citenamefont {Babaev}}]{zyuzin2017}%
  \BibitemOpen
  \bibfield  {author} {\bibinfo {author} {\bibfnamefont {A.}~\bibnamefont
  {Zyuzin}}, \bibinfo {author} {\bibfnamefont {J.}~\bibnamefont {Garaud}}, \
  and\ \bibinfo {author} {\bibfnamefont {E.}~\bibnamefont {Babaev}},\ }\href
  {https://arxiv.org/abs/1705.01718} {\bibinfo  {journal} {arXiv:1705.01718}\
  }\BibitemShut {NoStop}%
\bibitem [{\citenamefont {Yuan}\ \emph {et~al.}(2017)\citenamefont {Yuan},
  \citenamefont {He},\ and\ \citenamefont {Law}}]{Law2017}%
  \BibitemOpen
\bibfield  {journal} {  }\bibfield  {author} {\bibinfo {author} {\bibfnamefont
  {N.~F.~Q.}\ \bibnamefont {Yuan}}, \bibinfo {author} {\bibfnamefont {W.-Y.}\
  \bibnamefont {He}}, \ and\ \bibinfo {author} {\bibfnamefont {K.~T.}\
  \bibnamefont {Law}},\ }\href {\doibase 10.1103/PhysRevB.95.201109} {\bibfield
   {journal} {\bibinfo  {journal} {Phys. Rev. B}\ }\textbf {\bibinfo {volume}
  {95}},\ \bibinfo {pages} {201109} (\bibinfo {year} {2017})}\BibitemShut
  {NoStop}%
\bibitem [{\citenamefont {Chirolli}\ \emph {et~al.}(2017)\citenamefont
  {Chirolli}, \citenamefont {de~Juan},\ and\ \citenamefont
  {Guinea}}]{Chirolli2017}%
  \BibitemOpen
  \bibfield  {author} {\bibinfo {author} {\bibfnamefont {L.}~\bibnamefont
  {Chirolli}}, \bibinfo {author} {\bibfnamefont {F.}~\bibnamefont {de~Juan}}, \
  and\ \bibinfo {author} {\bibfnamefont {F.}~\bibnamefont {Guinea}},\ }\href
  {\doibase 10.1103/PhysRevB.95.201110} {\bibfield  {journal} {\bibinfo
  {journal} {Phys. Rev. B}\ }\textbf {\bibinfo {volume} {95}},\ \bibinfo
  {pages} {201110} (\bibinfo {year} {2017})}\BibitemShut {NoStop}%
\bibitem [{\citenamefont {Fu}(2016)}]{Fu2016}%
  \BibitemOpen
  \bibfield  {author} {\bibinfo {author} {\bibfnamefont {L.}~\bibnamefont
  {Fu}},\ }\href
  {http://www.nature.com/nphys/journal/v12/n9/full/nphys3793.html} {\bibfield
  {journal} {\bibinfo  {journal} {Nat. Phys.}\ }\textbf {\bibinfo {volume}
  {12}},\ \bibinfo {pages} {822} (\bibinfo {year} {2016})}\BibitemShut
  {NoStop}%
\bibitem [{\citenamefont {Behnia}(2017)}]{Behnia2017}%
  \BibitemOpen
  \bibfield  {author} {\bibinfo {author} {\bibfnamefont {K.}~\bibnamefont
  {Behnia}},\ }\href
  {https://www.nature.com/nphys/journal/v13/n2/full/nphys3932.html} {\bibfield
  {journal} {\bibinfo  {journal} {Nat. Phys.}\ }\textbf {\bibinfo {volume}
  {13}},\ \bibinfo {pages} {111} (\bibinfo {year} {2017})}\BibitemShut
  {NoStop}%
\bibitem [{\citenamefont {Fu}\ and\ \citenamefont {Berg}(2010)}]{Fu_Berg}%
  \BibitemOpen
  \bibfield  {author} {\bibinfo {author} {\bibfnamefont {L.}~\bibnamefont
  {Fu}}\ and\ \bibinfo {author} {\bibfnamefont {E.}~\bibnamefont {Berg}},\
  }\href {\doibase 10.1103/PhysRevLett.105.097001} {\bibfield  {journal}
  {\bibinfo  {journal} {Phys. Rev. Lett.}\ }\textbf {\bibinfo {volume} {105}},\
  \bibinfo {pages} {097001} (\bibinfo {year} {2010})}\BibitemShut {NoStop}%
\bibitem [{\citenamefont {Leggett}(1975)}]{Legget1975}%
  \BibitemOpen
  \bibfield  {author} {\bibinfo {author} {\bibfnamefont {A.~J.}\ \bibnamefont
  {Leggett}},\ }\href {\doibase 10.1103/RevModPhys.47.331} {\bibfield
  {journal} {\bibinfo  {journal} {Rev. Mod. Phys.}\ }\textbf {\bibinfo {volume}
  {47}},\ \bibinfo {pages} {331} (\bibinfo {year} {1975})}\BibitemShut
  {NoStop}%
\bibitem [{\citenamefont {Rice}\ and\ \citenamefont
  {Sigrist}(1995)}]{RiceSigrist}%
  \BibitemOpen
  \bibfield  {author} {\bibinfo {author} {\bibfnamefont {T.}~\bibnamefont
  {Rice}}\ and\ \bibinfo {author} {\bibfnamefont {M.}~\bibnamefont {Sigrist}},\
  }\href {http://iopscience.iop.org/article/10.1088/0953-8984/7/47/002/meta}
  {\bibfield  {journal} {\bibinfo  {journal} {Journal of Physics: Condensed
  Matter}\ }\textbf {\bibinfo {volume} {7}},\ \bibinfo {pages} {L643} (\bibinfo
  {year} {1995})}\BibitemShut {NoStop}%
\bibitem [{\citenamefont {Nomoto}\ and\ \citenamefont
  {Ikeda}(2016)}]{Nomoto2016}%
  \BibitemOpen
  \bibfield  {author} {\bibinfo {author} {\bibfnamefont {T.}~\bibnamefont
  {Nomoto}}\ and\ \bibinfo {author} {\bibfnamefont {H.}~\bibnamefont {Ikeda}},\
  }\href {\doibase 10.1103/PhysRevLett.117.217002} {\bibfield  {journal}
  {\bibinfo  {journal} {Phys. Rev. Lett.}\ }\textbf {\bibinfo {volume} {117}},\
  \bibinfo {pages} {217002} (\bibinfo {year} {2016})}\BibitemShut {NoStop}%
\bibitem [{\citenamefont {Kozii}\ and\ \citenamefont {Fu}(2015)}]{Kozii2015}%
  \BibitemOpen
  \bibfield  {author} {\bibinfo {author} {\bibfnamefont {V.}~\bibnamefont
  {Kozii}}\ and\ \bibinfo {author} {\bibfnamefont {L.}~\bibnamefont {Fu}},\
  }\href {\doibase 10.1103/PhysRevLett.115.207002} {\bibfield  {journal}
  {\bibinfo  {journal} {Phys. Rev. Lett.}\ }\textbf {\bibinfo {volume} {115}},\
  \bibinfo {pages} {207002} (\bibinfo {year} {2015})}\BibitemShut {NoStop}%
\bibitem [{\citenamefont {Wang}\ \emph {et~al.}(2016)\citenamefont {Wang},
  \citenamefont {Cho}, \citenamefont {Hughes},\ and\ \citenamefont
  {Fradkin}}]{Wang2016}%
  \BibitemOpen
  \bibfield  {author} {\bibinfo {author} {\bibfnamefont {Y.}~\bibnamefont
  {Wang}}, \bibinfo {author} {\bibfnamefont {G.~Y.}\ \bibnamefont {Cho}},
  \bibinfo {author} {\bibfnamefont {T.~L.}\ \bibnamefont {Hughes}}, \ and\
  \bibinfo {author} {\bibfnamefont {E.}~\bibnamefont {Fradkin}},\ }\href
  {\doibase 10.1103/PhysRevB.93.134512} {\bibfield  {journal} {\bibinfo
  {journal} {Phys. Rev. B}\ }\textbf {\bibinfo {volume} {93}},\ \bibinfo
  {pages} {134512} (\bibinfo {year} {2016})}\BibitemShut {NoStop}%
\bibitem [{\citenamefont {Ruhman}\ \emph {et~al.}(2017)\citenamefont {Ruhman},
  \citenamefont {Kozii},\ and\ \citenamefont {Fu}}]{Ruhman2017}%
  \BibitemOpen
  \bibfield  {author} {\bibinfo {author} {\bibfnamefont {J.}~\bibnamefont
  {Ruhman}}, \bibinfo {author} {\bibfnamefont {V.}~\bibnamefont {Kozii}}, \
  and\ \bibinfo {author} {\bibfnamefont {L.}~\bibnamefont {Fu}},\ }\href
  {\doibase 10.1103/PhysRevLett.118.227001} {\bibfield  {journal} {\bibinfo
  {journal} {Phys. Rev. Lett.}\ }\textbf {\bibinfo {volume} {118}},\ \bibinfo
  {pages} {227001} (\bibinfo {year} {2017})}\BibitemShut {NoStop}%
\bibitem [{\citenamefont {Venderbos}\ \emph
  {et~al.}(2016{\natexlab{b}})\citenamefont {Venderbos}, \citenamefont
  {Kozii},\ and\ \citenamefont {Fu}}]{Venderbos2016_pairing}%
  \BibitemOpen
  \bibfield  {author} {\bibinfo {author} {\bibfnamefont {J.~W.~F.}\
  \bibnamefont {Venderbos}}, \bibinfo {author} {\bibfnamefont {V.}~\bibnamefont
  {Kozii}}, \ and\ \bibinfo {author} {\bibfnamefont {L.}~\bibnamefont {Fu}},\
  }\href {\doibase 10.1103/PhysRevB.94.180504} {\bibfield  {journal} {\bibinfo
  {journal} {Phys. Rev. B}\ }\textbf {\bibinfo {volume} {94}},\ \bibinfo
  {pages} {180504} (\bibinfo {year} {2016}{\natexlab{b}})}\BibitemShut
  {NoStop}%
\bibitem [{\citenamefont {Fu}(2015)}]{Fu2015}%
  \BibitemOpen
  \bibfield  {author} {\bibinfo {author} {\bibfnamefont {L.}~\bibnamefont
  {Fu}},\ }\href {\doibase 10.1103/PhysRevLett.115.026401} {\bibfield
  {journal} {\bibinfo  {journal} {Phys. Rev. Lett.}\ }\textbf {\bibinfo
  {volume} {115}},\ \bibinfo {pages} {026401} (\bibinfo {year}
  {2015})}\BibitemShut {NoStop}%
\bibitem [{\citenamefont {Harter}\ \emph {et~al.}(2017)\citenamefont {Harter},
  \citenamefont {Zhao}, \citenamefont {Yan}, \citenamefont {Mandrus},\ and\
  \citenamefont {Hsieh}}]{Harter2017}%
  \BibitemOpen
  \bibfield  {author} {\bibinfo {author} {\bibfnamefont {J.}~\bibnamefont
  {Harter}}, \bibinfo {author} {\bibfnamefont {Z.}~\bibnamefont {Zhao}},
  \bibinfo {author} {\bibfnamefont {J.-Q.}\ \bibnamefont {Yan}}, \bibinfo
  {author} {\bibfnamefont {D.}~\bibnamefont {Mandrus}}, \ and\ \bibinfo
  {author} {\bibfnamefont {D.}~\bibnamefont {Hsieh}},\ }\href
  {http://science.sciencemag.org/content/356/6335/295} {\bibfield  {journal}
  {\bibinfo  {journal} {Science}\ }\textbf {\bibinfo {volume} {356}},\ \bibinfo
  {pages} {295} (\bibinfo {year} {2017})}\BibitemShut {NoStop}%
\bibitem [{\citenamefont {Wray}\ \emph {et~al.}(2010)\citenamefont {Wray},
  \citenamefont {Xu}, \citenamefont {Xia}, \citenamefont {San~Hor},
  \citenamefont {Qian}, \citenamefont {Fedorov}, \citenamefont {Lin},
  \citenamefont {Bansil}, \citenamefont {Cava},\ and\ \citenamefont
  {Hasan}}]{Hasan2010}%
  \BibitemOpen
  \bibfield  {author} {\bibinfo {author} {\bibfnamefont {L.~A.}\ \bibnamefont
  {Wray}}, \bibinfo {author} {\bibfnamefont {S.-Y.}\ \bibnamefont {Xu}},
  \bibinfo {author} {\bibfnamefont {Y.}~\bibnamefont {Xia}}, \bibinfo {author}
  {\bibfnamefont {Y.}~\bibnamefont {San~Hor}}, \bibinfo {author} {\bibfnamefont
  {D.}~\bibnamefont {Qian}}, \bibinfo {author} {\bibfnamefont {A.~V.}\
  \bibnamefont {Fedorov}}, \bibinfo {author} {\bibfnamefont {H.}~\bibnamefont
  {Lin}}, \bibinfo {author} {\bibfnamefont {A.}~\bibnamefont {Bansil}},
  \bibinfo {author} {\bibfnamefont {R.~J.}\ \bibnamefont {Cava}}, \ and\
  \bibinfo {author} {\bibfnamefont {M.~Z.}\ \bibnamefont {Hasan}},\ }\href
  {https://www.nature.com/nphys/journal/v6/n11/abs/nphys1762.html} {\bibfield
  {journal} {\bibinfo  {journal} {Nat. Phys.}\ }\textbf {\bibinfo {volume}
  {6}},\ \bibinfo {pages} {855} (\bibinfo {year} {2010})}\BibitemShut {NoStop}%
\bibitem [{\citenamefont {Lawson}\ \emph {et~al.}(2012)\citenamefont {Lawson},
  \citenamefont {Hor},\ and\ \citenamefont {Li}}]{Lawson2012}%
  \BibitemOpen
  \bibfield  {author} {\bibinfo {author} {\bibfnamefont {B.~J.}\ \bibnamefont
  {Lawson}}, \bibinfo {author} {\bibfnamefont {Y.~S.}\ \bibnamefont {Hor}}, \
  and\ \bibinfo {author} {\bibfnamefont {L.}~\bibnamefont {Li}},\ }\href
  {\doibase 10.1103/PhysRevLett.109.226406} {\bibfield  {journal} {\bibinfo
  {journal} {Phys. Rev. Lett.}\ }\textbf {\bibinfo {volume} {109}},\ \bibinfo
  {pages} {226406} (\bibinfo {year} {2012})}\BibitemShut {NoStop}%
\bibitem [{\citenamefont {Qi}\ \emph {et~al.}(2008)\citenamefont {Qi},
  \citenamefont {Hughes},\ and\ \citenamefont {Zhang}}]{Qi2008}%
  \BibitemOpen
  \bibfield  {author} {\bibinfo {author} {\bibfnamefont {X.-L.}\ \bibnamefont
  {Qi}}, \bibinfo {author} {\bibfnamefont {T.~L.}\ \bibnamefont {Hughes}}, \
  and\ \bibinfo {author} {\bibfnamefont {S.-C.}\ \bibnamefont {Zhang}},\ }\href
  {\doibase 10.1103/PhysRevB.78.195424} {\bibfield  {journal} {\bibinfo
  {journal} {Phys. Rev. B}\ }\textbf {\bibinfo {volume} {78}},\ \bibinfo
  {pages} {195424} (\bibinfo {year} {2008})}\BibitemShut {NoStop}%
\bibitem [{\citenamefont {Goswami}\ and\ \citenamefont {Roy}(2014)}]{Roy2014}%
  \BibitemOpen
  \bibfield  {author} {\bibinfo {author} {\bibfnamefont {P.}~\bibnamefont
  {Goswami}}\ and\ \bibinfo {author} {\bibfnamefont {B.}~\bibnamefont {Roy}},\
  }\href {\doibase 10.1103/PhysRevB.90.041301} {\bibfield  {journal} {\bibinfo
  {journal} {Phys. Rev. B}\ }\textbf {\bibinfo {volume} {90}},\ \bibinfo
  {pages} {041301} (\bibinfo {year} {2014})}\BibitemShut {NoStop}%
\bibitem [{\citenamefont {Wang}\ and\ \citenamefont {Fu}()}]{Wang_Fu}%
  \BibitemOpen
  \bibfield  {author} {\bibinfo {author} {\bibfnamefont {Y.}~\bibnamefont
  {Wang}}\ and\ \bibinfo {author} {\bibfnamefont {L.}~\bibnamefont {Fu}},\
  }\href {https://arxiv.org/abs/1703.06880} {\bibinfo  {journal}
  {arXiv:1703.06880}\ }\BibitemShut {NoStop}%
\bibitem [{\citenamefont {Kozii}\ \emph {et~al.}(2016)\citenamefont {Kozii},
  \citenamefont {Venderbos},\ and\ \citenamefont {Fu}}]{kozii2016}%
  \BibitemOpen
\bibfield  {journal} {  }\bibfield  {author} {\bibinfo {author} {\bibfnamefont
  {V.}~\bibnamefont {Kozii}}, \bibinfo {author} {\bibfnamefont {J.~W.}\
  \bibnamefont {Venderbos}}, \ and\ \bibinfo {author} {\bibfnamefont
  {L.}~\bibnamefont {Fu}},\ }\href
  {http://advances.sciencemag.org/content/2/12/e1601835} {\bibfield  {journal}
  {\bibinfo  {journal} {Sci. Adv.}\ }\textbf {\bibinfo {volume} {2}},\ \bibinfo
  {pages} {e1601835} (\bibinfo {year} {2016})}\BibitemShut {NoStop}%
\bibitem [{\citenamefont {Brydon}\ \emph {et~al.}(2014)\citenamefont {Brydon},
  \citenamefont {Das~Sarma}, \citenamefont {Hui},\ and\ \citenamefont
  {Sau}}]{Brydon2014}%
  \BibitemOpen
  \bibfield  {author} {\bibinfo {author} {\bibfnamefont {P.~M.~R.}\
  \bibnamefont {Brydon}}, \bibinfo {author} {\bibfnamefont {S.}~\bibnamefont
  {Das~Sarma}}, \bibinfo {author} {\bibfnamefont {H.-Y.}\ \bibnamefont {Hui}},
  \ and\ \bibinfo {author} {\bibfnamefont {J.~D.}\ \bibnamefont {Sau}},\ }\href
  {\doibase 10.1103/PhysRevB.90.184512} {\bibfield  {journal} {\bibinfo
  {journal} {Phys. Rev. B}\ }\textbf {\bibinfo {volume} {90}},\ \bibinfo
  {pages} {184512} (\bibinfo {year} {2014})}\BibitemShut {NoStop}%
\bibitem [{\citenamefont {Wan}\ and\ \citenamefont {Savrasov}(2014)}]{Wan2014}%
  \BibitemOpen
  \bibfield  {author} {\bibinfo {author} {\bibfnamefont {X.}~\bibnamefont
  {Wan}}\ and\ \bibinfo {author} {\bibfnamefont {S.~Y.}\ \bibnamefont
  {Savrasov}},\ }\href {https://www.nature.com/articles/ncomms5144} {\bibfield
  {journal} {\bibinfo  {journal} {Nat. Commun.}\ }\textbf {\bibinfo {volume}
  {5}},\ \bibinfo {pages} {4144} (\bibinfo {year} {2014})}\BibitemShut
  {NoStop}%
\bibitem [{\citenamefont {Zhang}\ \emph {et~al.}(2009)\citenamefont {Zhang},
  \citenamefont {Liu}, \citenamefont {Qi}, \citenamefont {Dai}, \citenamefont
  {Fang},\ and\ \citenamefont {Zhang}}]{Zhang2009}%
  \BibitemOpen
  \bibfield  {author} {\bibinfo {author} {\bibfnamefont {H.}~\bibnamefont
  {Zhang}}, \bibinfo {author} {\bibfnamefont {C.-X.}\ \bibnamefont {Liu}},
  \bibinfo {author} {\bibfnamefont {X.-L.}\ \bibnamefont {Qi}}, \bibinfo
  {author} {\bibfnamefont {X.}~\bibnamefont {Dai}}, \bibinfo {author}
  {\bibfnamefont {Z.}~\bibnamefont {Fang}}, \ and\ \bibinfo {author}
  {\bibfnamefont {S.-C.}\ \bibnamefont {Zhang}},\ }\href
  {https://www.nature.com/nphys/journal/v5/n6/abs/nphys1270.html} {\bibfield
  {journal} {\bibinfo  {journal} {Nat. Phys.}\ }\textbf {\bibinfo {volume}
  {5}},\ \bibinfo {pages} {438} (\bibinfo {year} {2009})}\BibitemShut {NoStop}%
\bibitem [{\citenamefont {Wang}\ and\ \citenamefont
  {Zhang}(2012)}]{Wang2012Phonon}%
  \BibitemOpen
  \bibfield  {author} {\bibinfo {author} {\bibfnamefont {B.-T.}\ \bibnamefont
  {Wang}}\ and\ \bibinfo {author} {\bibfnamefont {P.}~\bibnamefont {Zhang}},\
  }\href {http://aip.scitation.org/doi/abs/10.1063/1.3689759?journalCode=apl}
  {\bibfield  {journal} {\bibinfo  {journal} {Appl. Phys. Lett.}\ }\textbf
  {\bibinfo {volume} {100}},\ \bibinfo {pages} {082109} (\bibinfo {year}
  {2012})}\BibitemShut {NoStop}%
\end{thebibliography}%

\end{document}